\documentclass[fleqn,usenatbib]{mnras}

\usepackage{newtxtext,newtxmath}

\usepackage[T1]{fontenc}

\DeclareRobustCommand{\VAN}[3]{#2}
\let\VANthebibliography\thebibliography
\def\thebibliography{\DeclareRobustCommand{\VAN}[3]{##3}\VANthebibliography}

\usepackage{graphicx}	
\usepackage{amsmath}	

\newcommand{\kms}{km\,s$^{-1}$}

\title[Methanol emission in the  $J_1 - J_0$ A$^{-+}$ line series]{The methanol emission in the  $J_1 - J_0$ A$^{-+}$ line series as a tracer of specific physical conditions in high-mass star-forming regions}

\author[S. V. Salii et al.]
{
Svetlana V. Salii,$^{1}$\thanks{E-mail: svetlana.salii@urfu.ru (SVS)}
Igor I. Zinchenko,$^{2}$ 
Sheng-Yuan Liu,$^{3}$  
Andrej M. Sobolev,$^{1}$ 
\newauthor 
Artis  Aberfelds$^{4}$ and Yu-Nung Su$^{3}$  \\
$^{1}$Astronomical Observatory, Ural Federal University, 51, Lenin ave.,  Ekaterinburg, Russia\\
$^{2}$Institute of Applied Physics of the Russian Academy of Sciences, 46 Ul'yanov~str., 603950 Nizhny Novgorod, Russia.\\
$^{3}$Institute of Astronomy and Astrophysics, Academia Sinica, 11F of ASMAB, AS/NTU No.1, Sec. 4, Roosevelt Rd, Taipei 10617, Taiwan\\
$^{4}$Engineering Research Institute ``Ventspils International Radio Astronomy Center'', Ventspils University of Applied Sciences, Inzenieru Str. 101, Ventspils,\\ LV-3601, Latvia\\
}

\date{Accepted XXX. Received YYY; in original form ZZZ}

\pubyear{2022}

\begin{document}
\label{firstpage}
\pagerange{\pageref{firstpage}--\pageref{lastpage}}
\maketitle

\begin{abstract}
We present results of the investigations of the properties of the methanol $J_1 - J_0$ A$^{-+}$ line series motivated by the recent serendipitous detection of the maser emission in the $14_1 - 14_0$ A$^{-+}$ line at 349~GHz in S255IR-SMA1 soon after the accretion burst. The study includes further observations of several lines of this series in S255IR with the SMA, a mini-survey   of methanol lines in the 0.8 mm range toward a sample  of bright 6.7~GHz methanol maser sources with the IRAM-30m telescope, and theoretical modeling. We found that the maser component of the $14_1 - 14_0$ A$^{-+}$ line in S255IR decayed   by more than order of magnitude in comparison with that in 2016.   
 No clear sign of maser emission is observed in other lines of this series in the SMA observations except the $7_1 - 7_0$ A$^{-+}$ line where an additional bright component is detected at the velocity of the maser emission observed earlier in the $14_1 - 14_0$ A$^{-+}$ line. Our LVG model constrains the ranges of the physical parameters that matches the observed emission intensities. 
No   obvious  maser emission in the $J_1 - J_0$ A$^{-+}$ lines was detected in the mini-survey of the 6.7~GHz methanol maser sources, though one component in NGC7538 may represent a weak maser.
In general, the maser effect in the $J_1 - J_0$ A$^{-+}$ lines may serve as a tracer of rather hot environments and in particular luminosity flaring events during high mass star formation.

\end{abstract}




\section{Introduction} \label{sect:intro}



Methanol, CH$_3$OH, is an important component of the interstellar gas in regions of star formation. 
  Methanol is known as a good tracer of the physical conditions of the molecular gas in high-mass star-forming regions  \citep{2004ApJ...609..231S,2006ARep...50..965S,Cragg_et_al_2005_MNRAS_360}. Due to its molecular structural properties,  CH$_3$OH  has a rich radio spectrum. 
 The intensities and their ratios for the lines produced in different series of transitions depend on the physical conditions in molecular clouds and can be used for the estimations of the physical parameters \citep{1992AZh....69.1148S,2002ARep...46..955S,2006ARep...50..965S,2006MNRAS.373..411V,2004A&A...422..573L,2007A&A...466..215L}.  
Under certain  physical conditions in high-mass star-forming regions a maser amplification can occur. 
Analysis of observational data has shown that there are two classes   of  methanol masers, Class I and
Class II, which emit in different transition sets \citep{1991ASPC...16..119M}. Class II methanol masers reside closer to young stellar objects (YSOs) and trace circumstellar disks and inner parts of the outflows, while Class I masers are associated with more distant parts of the outflows and shocked regions   \citep{2007IAUS..242...81S}. 
 
  Many methanol masering transitions have been observed.
More than 20 Class I methanol maser lines \citep[][and the references therein]{Ladeyschikov_et_al_2019}  are observed nowadays. 
Nearly 30 among about 100 theoretically predicted Class II methanol maser transitions \citep{Cragg_et_al_2005_MNRAS_360} are listed  as observed  in the  database MaserDB\footnote{\url{http://maserdb.net/search.pl} \citep{Ladeyschikov_et_al_2019} }.    
 New maser transitions are still getting discovered,  e.g. 22 new Class II methanol masers were discovered in G358.931-0.030 in 2019  \citep{2019ApJ...881L..39B,2019ApJ...876L..25B,2019MNRAS.489.3981M}. It is noteworthy that only 5 of those were listed as bright masers in the theoretical models \citep{Cragg_et_al_2005_MNRAS_360}.  
  Furthermore,  a very  bright line, 
  $\sim 4000$~K ($\sim 6$~Jy/beam with the $0{\farcs}10\times0{\farcs}15$ beam) at 349.1 GHz,
  detected   in the high-mass star forming region S255IR in 2016 with ALMA, was identified   as  the CH$_3$OH $14_1 - 14_0$ A$^{-+}$ transition and was qualitatively classified as an unpredicted Class II maser \citep{Zinchenko_et_al_2017_AA_606}. Remarkably, the transition mentioned above has never been observed or predicted to be masering \citep{Sobolev_et_al_1997_MNRAS_288, Cragg_et_al_2005_MNRAS_360, Voronkov_et_al_2012_IAUS_287}. 

 The maser emission in the  $14_1 - 14_0$ A$^{-+}$  line  
was detected in S255IR SMA1 about 1 year after the 
accretion burst event seen in the IR band by \citet{Caratti17}. 
Subsequent observations in 2017   showed  a significant (by about 40\%) decay of this maser line, in about the same amount as the decay of the submillimeter continuum emission  \citep{Liu2018}. These data indicate that this maser emission can be related to the luminosity bursts during the process of high-mass star formation and raise the questions   such as  whether it can be observed in the other lines of this series, how frequent this phenomenon is in similar objects, and  which  physical conditions are required for  the  excitation of such masers. 

In order to answer these questions, we performed further observations of S255IR in several methanol lines of the $J_1 - J_0$ A$^{-+}$ series at a high angular resolution and conducted a survey toward a sample of well-known strong Class II methanol maser sources in the 349~GHz line and simultaneously in several transitions of the same series.
 The state of activity of associated young stellar objects was checked by the 6.7 GHz methanol maser monitoring.  
Then, we performed theoretical modeling of the excitation of the corresponding transitions. The results are presented below.

\section{Observations}
\label{sect:Obs}

\subsection{S255IR SMA Observations}

The observations toward S255IR were carried out with the SMA on 2019 March 21 with the array in its extended configuration.
  The phase center was set at J2000 ${\rm R.A.}= 06^h 12^m 54{\fs}015$ and ${\rm Dec.} = 17^\circ 59'23{\farcs}05$. 
The dual receiver observing mode was employed with the 345~GHz receiver centered at 302.97~GHz and the 400~GHz receiver centered at 353.86~GHz.
The half-power width of the SMA primary beam is about 36$"$ at 345 GHz.

The SWARM (SMA Wideband Astronomical ROACH2 Machine) correlator   enabled  a simultaneous spectral coverage of 32 GHz in total (290.97 -- 298.97~GHz, 306.97 -- 314.97~GHz, 341.86 -- 349.86~GHz, 357.86 -- 365.86~GHz) and an uniform spectral resolution  
  for all bands at a channel width of 140 kHz. 
The combination of the dual frequency setup and the backend correlator allowed us to cover six methanol $J_1 - J_0~A^{-+}$ transitions as listed in Table~\ref{tab:J1_J0Amp_SMAobs_list}.
3C279 served as the bandpass calibrator and asteroid Pallas was used for the absolute flux calibration (with its flux ranging from 1.88 Jy at 295 GHz to 2.7~Jy  at 360 GHz).
The nearby compact radio sources 0854+201 ($\sim$ 2.0 Jy) and 0510+180 ($\sim$ 1.8 Jy) were employed as the complex gain calibrators. 
Typical absolute flux density is estimated to have an uncertainty of $\sim$ 20\%.

We calibrated the data using the MIR package  \citep{Scoville93}  and  imaged the data using the MIRIAD software   \citep{Sault95}.
The projected baselines range from about 30~m to 210~m, leading to an angular resolution of $\sim$ {1{\farcs}0} $\times$ {0{\farcs}8} at 307~GHz and $\sim$ {0{\farcs}9} $\times$ {0{\farcs}6} at 349~GHz under  robust weighting. 
The resulting molecular line brightness sensitivity is {0.2} Jy~beam$^{-1}$ or equivalently $\sim$ {3}~K for a spectral resolution of {0.25}~\kms.
\begin{table} 
\caption{Methanol (CH$_3$OH) $J_{1} - J_{0}$ A$^{- +}$ lines observed with the SMA   toward  S255IR} 
\label{tab:J1_J0Amp_SMAobs_list}
\centering
\begin{tabular}{c c r }
\hline \hline
Rest Frequency (GHz) & Quantum Numbers & E$_{u}$ (K)  
\\
\hline
	307.16594 &	$4(1,3)-4(0,4)$ A$^{- +}$		& 38.0  
	\\
	309.29040 &	$5(1,4)-5(0,5)$ A$^{- +}$		& 49.7  
	\\
	311.85264 &	$6(1,5)-6(0,6)$ A$^{- +}$		& 63.7  
	\\
	314.85955 &	$7(1,6)-7(0,7)$ A$^{- +}$		& 80.1  
	\\
	342.72983 &	$13(1,12)-13(0,13)$ A$^{- +}$	& 227.5  
	\\
	349.10702 &	$14(1,13)-14(0,14)$ A$^{- +}$	& 260.2  
	\\
\hline 
\end{tabular}
\end{table}

\subsection{IRAM-30m observations}
  A total of 15  brightest 6.7~GHz Class II methanol maser sources were selected for our survey observations, which were performed with the 30m IRAM radio telescope in January 2019 (see Table~\ref{tab:sources_list}).  
  We selected the objects with their flux densities at 6.7~GHz $F >1000$~Jy according to the database of astrophysical masers (MaserDb)\footnote{\url{https://maserdb.net}, \citep{Ladeyschikov_et_al_2019}} 
 and at sufficiently high elevation at the 30m IRAM site.

The observations were performed in the wobbler switching mode using the  EMIR receiver in four frequency bands:  326.7--330.7, 329.5--334.5,  342.4--346.4 and  346.1--350.2~GHz. At these frequencies the antenna HPBW and the spectral resolution are $7.5''$ and $\sim 1.4$~\kms, respectively.  The $1\sigma$ rms for different sources varied from 0.05 to 0.5 K in the 326.7--330.7 band and from 0.02 to 0.05 K in other bands. 
 
 The antenna temperature calibration was made by the standard
chopper-wheel method. The data reduction was performed with the GILDAS package\footnote{\url{http://www.iram.fr/IRAMFR/GILDAS}}.

\begin{table*}
	\centering
	\caption{  Young stellar objects associated with the brightest 6.7 GHz Class II methanol masers that were included in the observations.  
	 In columns 6 and 7, $V_{0}$ and $F$ are, respectively, the Radial Velocity and Flux density of the strongest maser spot according to the database of astrophysical masers (MaserDb) \citep{Ladeyschikov_et_al_2019}.
	 The values of $V_{l}$ and $V_{h}$ indicate the broadest interval of the maser emission according to MaserDb.   
	 The facilities (IRAM, RT-32 and SMA) that were employed in our 2019 observations are labeled by the plus sign in the last three columns. 
	} \label{tab:sources_list}
\begin{tabular}{clccrrrccc}
\hline 
N&Source & RA(2000)& DEC(2000)& $V_{l};V_{h}$ &$V_{0}$&$  F$&\multicolumn{3}{c}{Observations  2019} \\ 
&       & $\,^h\,:\,^m\,:\,^s\,$   & $\,^\circ:\,':\,''$ & \kms &\kms& Jy&IRAM&RT-32&SMA\\ 
\hline 
1&G008.831$-0.028$                & 18:05:25.666& $-21$:19:25.48& $-6 ; 6^c$    & $-0.51^a$ &  2096.42$^a$&+&&\\ 
2&G009.621$+0.195$                & 18:06:14.662& $-20$:31:31.47& $-4 ; 9^d$   & $  1.43^a$ & 13056.89$^a$&+&&\\ 
3&G012.680$-0.182$ (W33B)         & 18:13:54.741& $-18$:01:47.19&  $50 ; 61^d$  & $55.63^a$ &  5519.22$^a$&+&&\\ 
4&G012.908$+0.260$ (W33A)         & 18:14:39.527& $-17$:52:00.59&  $35 ; 47^d$  & $38.00^a$ &  1764.15$^a$&+&&\\ 
5&G023.009$-0.410$                & 18:34:40.281& $-09$:00:38.23&  $68 ; 87.5^e$  & $76.31^a$ &  4523.72$^a$&+&&\\ 
6&G025.709$+0.043$ (W98)          & 18:38:03.138& $-06$:24:15.32&  $89 ; 103.5^e$ & $92.70^a$ &  2635.38$^a$&+&&\\ 
7&G035.200$-1.736$ (W48)          & 19:01:45.548& $+01$:13:32.88&  $39 ; 47^d$ & $42.67^a$ &  1061.27$^a$&+&&\\ 
8&G037.429$+1.517$                & 18:54:14.238& $+04$:41:41.00&  $40.2 ; 52.7^e$  & $43.12^a$ &  1206.47$^a$&+&&\\ 
9&G049.489$-0.387$ (W51e2)                & 19:23:43.938& $+14$:30:34.09&  $49.9 ; 65^f$  & $55.41^a$ &  2862.55$^a$&+&+&\\ 
10&G081.871$+0.780$ (W75N)        & 20:38:36.425& $+42$:37:34.56&  $0 ; 13^g$   &  $6.32^a$  &  1865.45$^a$&+&+&\\ 
11&G109.870$+2.114$ (Cepheus A)   & 22:56:17.883& $+62$:01:49.53&  $-6 ; -1^h$& $-4.04^a$ &  2133.03$^a$&+&+&\\ 
12&G111.542$+0.776$ (NGC 7538C)   & 23:13:45.386& $+61$:28:09.81&  $-62 ; -48^g$&$-58.56^a$ &  2331.18$^a$&+&+&\\ 
13&G133.947$+1.064$ (W3(OH))      & 02:27:03.810& $+61$:52:25.18&  $-48 ; -41^g$ &$-44.86^a$ & 37474.84$^a$&+&+&\\ 
14&G188.946$+0.886$ (AFGL5180)    & 06:08:53.343& $+21$:38:29.14&  $-4 ; 12^d$ & $ 9.92^a$ &  2741.29$^a$&+&&\\ 
15&G192.600$-0.048$ (S255IR)      & 06:12:54.010& $+17$:59:23.06&  $2 ; 8^b$  &  $5.6^b$ &   1600.00$^b$&+&+&+\\ 
\hline
\multicolumn{10}{l}{\parbox{0.9\linewidth}{
$^a$\citet{Hu_et_al_2016}, 
$^b$\citet{2015ATel.8286....1F}, 
$^c$\citet{2009A&A...507.1117X}, 
$^d$\citet{2009PASA...26..454C}, 
$^e$\citet{2015MNRAS.450.4109B}, 
$^f$\citet{2007ApJ...656..255P},  
$^g$\citet{2005A&A...432..737P}, 
$^h$\citep{1991ApJ...380L..75M}}
}
\end{tabular}
\end{table*}

\subsection{Observations at the RT-32 of the Ventspils International Radio Astronomy Centre}
At about the same time in January 2019,   a  majority (6 out of 7) of the sources with declination greater than 14 degrees    (see Tab.~\ref{tab:sources_list})  
were observed at the Ventspils International Radio Astronomy Centre at 6.7~GHz  using the Irbene 32 meter fully steerable radio telescope RT-32 with a cryogenic receiver at 6.7 GHz.
 \,The Rohde \& Swartz FSW43 spectrometer was used as the back-end to collect the spectral data. 
Standard frequency switch was used \citep{Wilson} with 30~sec long switching steps and 30~min total on-source time. The back-end was configured to 2~MHz  band with 4096 channels.

\section{Results}
\label{sect:results}

\subsection{Methanol line emission in S255IR observed with SMA} \label{sec:sma_res}
\begin{figure}
    \centering \includegraphics[width=0.93\columnwidth]{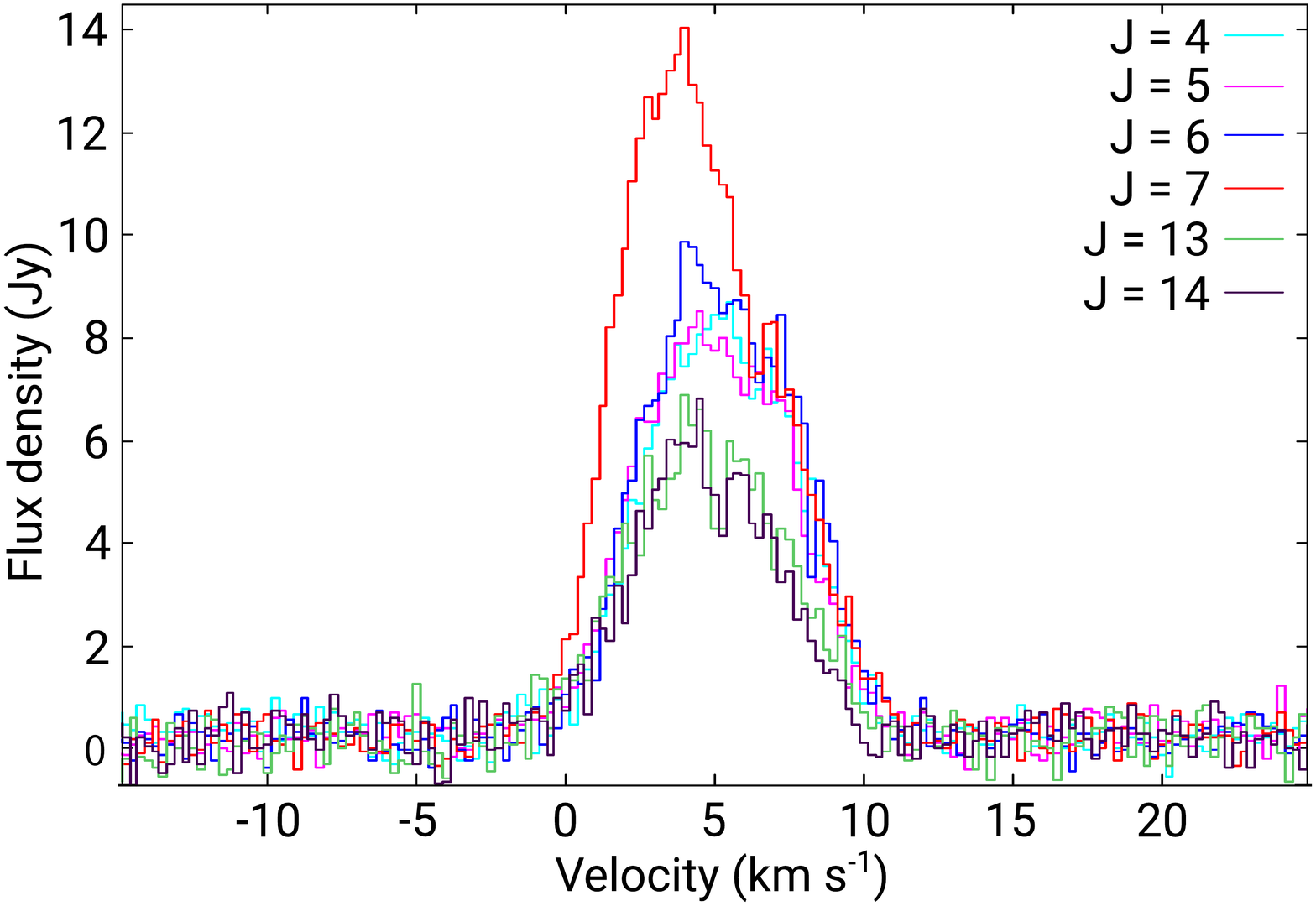}
	\includegraphics[width=\columnwidth]{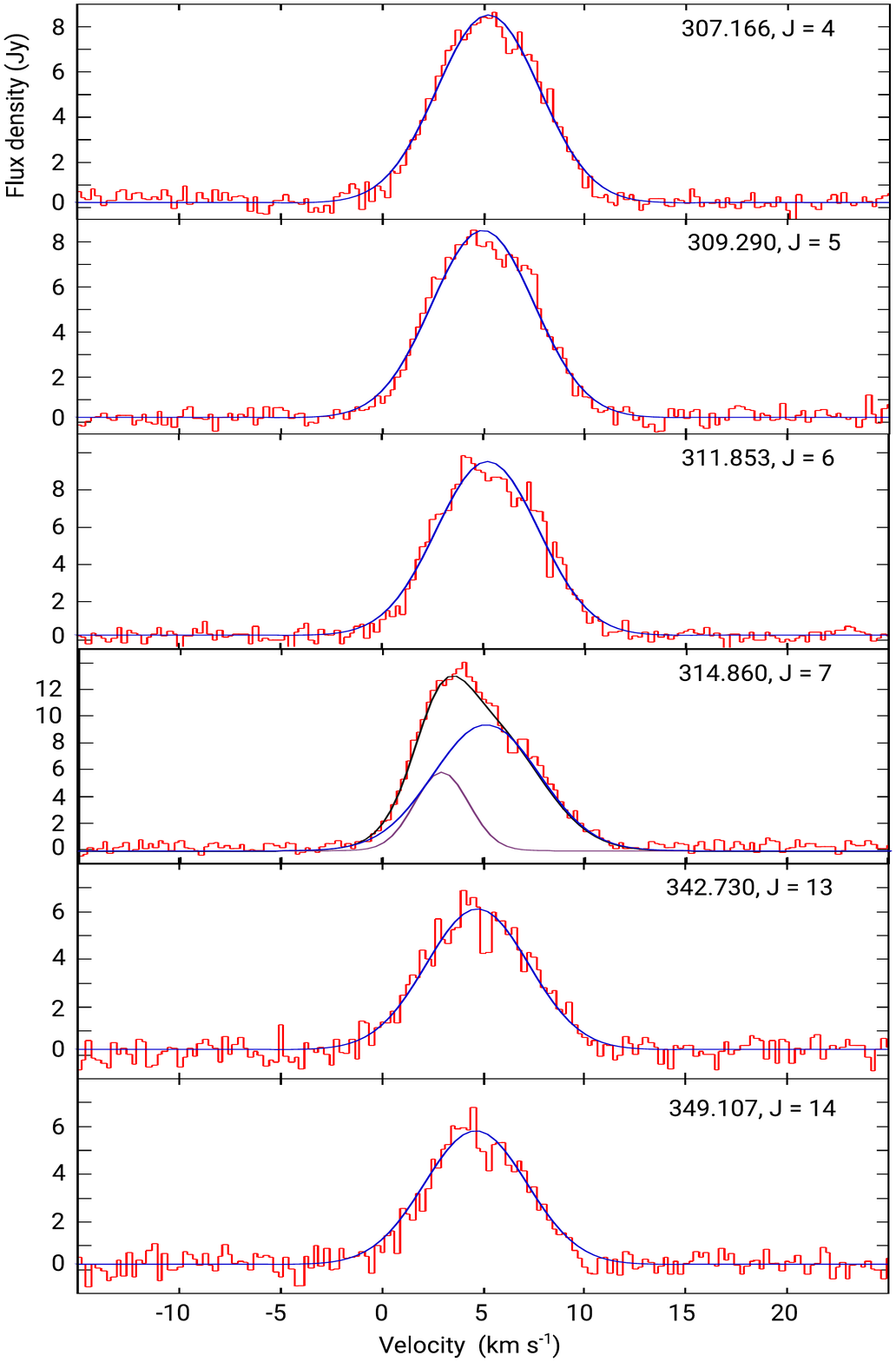}
    \caption{ Top panel: An overlay of the spectra of the $J_1 - J_0$ A$^{-+}$ CH$_3$OH line series observed with the SMA toward S255~IR. Bottom panels: Individual spectra of the $J_1 - J_0$ A$^{-+}$ CH$_3$OH line series observed with the SMA toward S255~IR. The single Gaussian fits with the peak at $\sim5$~\kms\ are overlaid in blue. The two Gaussian fits with peaks at 2.8~\kms\ and 5~\kms\ are plotted in violet and blue, respectively.   The rest frequencies in GHz and quantum numbers of the line series are given in the upper right corners. 
    }  \label{fig:J1_J0_Amp_sp_SMA}
\end{figure}

The SMA data include the methanol lines of the $J_1 - J_0$ A$^{-+}$ series with $J=4-7, 13, 14$. In Fig.~\ref{fig:J1_J0_Amp_sp_SMA},   we present  in the upper image  a superposition of spectra of these lines integrated over a circle of 3$^{\prime\prime}$ in diameter. The signal to noise ratio  in these data exceeds 100, which
is good enough for a detailed study of the line profiles.  

Five spectra from six are reasonably well fitted by a single Gaussian with a peak at about $5\pm0.05$~\kms\ (  see the bottom panels of Fig.~\ref{fig:J1_J0_Amp_sp_SMA} ), which corresponds to the systemic velocity of the core \citep{2015ApJ...810...10Z, Liu2020}.   The amplitude of this component is comparable for all transitions under consideration.    The full width at half maximum (FWHM)   of this Gaussian is about 6~\kms\ (with the uncertainties $\sim$0.1~\kms). 

Clearly in the lines with $J= 7$ and possibly in those with $J= 6$ an additional peak at about 2.5~\kms\ can be distinguished. 
We fitted this additional component  by a Gaussian with   a  central velocity of $2.8\pm0.1$~\kms\ and   FWHM  $3.0\pm0.1$~\kms. 
It coincides with the maser feature reported by \citet{Zinchenko_et_al_2017_AA_606}. 
  The flux density of this component is  $5.8\pm0.2$~Jy,   whereas it is practically absent in the other lines of this series (see Fig.~\ref{fig:J1_J0_Amp_sp_SMA}). 
  We could not find any reasonable identification for this component with lines of other molecules. 

Remarkably, the additional component is not detected confidently in the methanol  $14_1 - 14_0$ A$^{-+}$ line at 349.1~GHz in the present observations. The amplitude of such component in this line is   not  more than $\sim$1~Jy, which is about 20 times weaker than that in 2016.

\begin{figure}
	\includegraphics[width=0.49\textwidth]{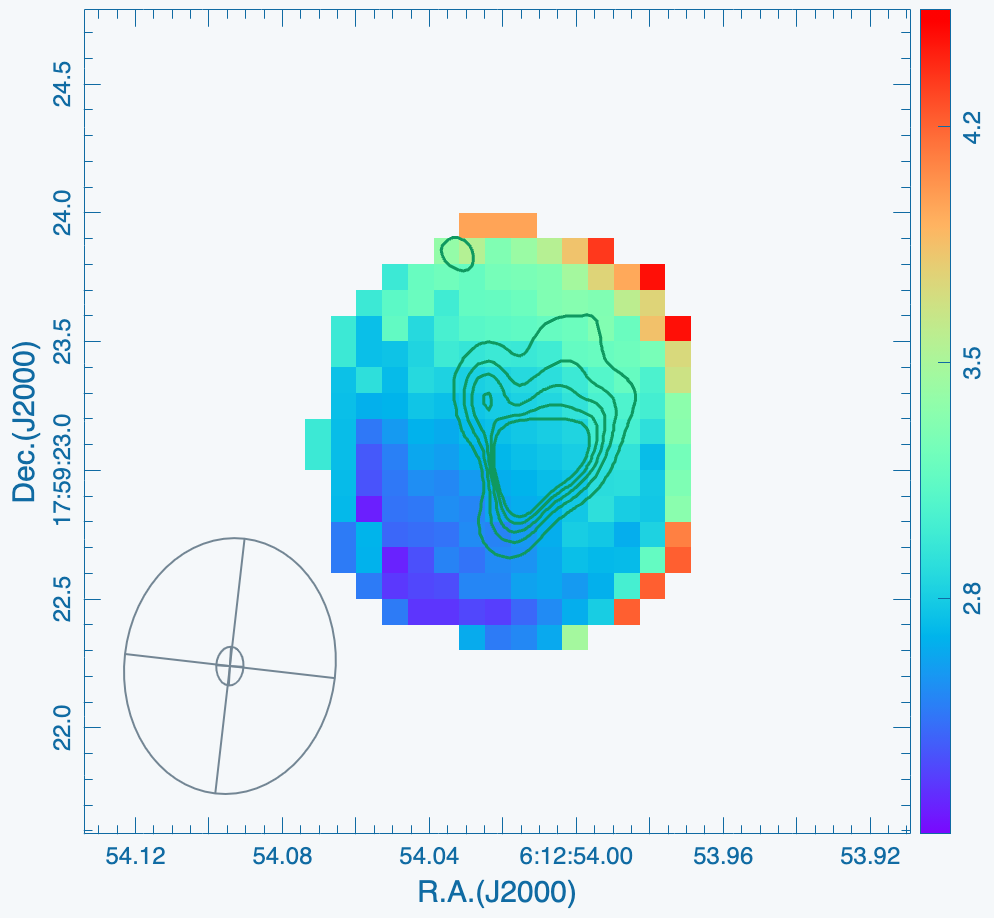}
    \caption{ The first-moment map of the CH$_3$OH emission in the 2.8~\kms\ component of the $7_1 - 7_0$ A$^{-+}$ line at 314.860 GHz observed with the SMA and overlaid with contours of the 0.9~mm continuum emission measured with ALMA \citep{Zinchenko_et_al_2017_AA_606,Liu2018,Liu2020}.  The color scale corresponds to the velocity in \kms. The contour levels are (1.15, 2.08, 3.01, 3.93, 4.86, 13)$\times$10$^{-2}$~Jy\,beam$^{-1}$. The synthesized beams are shown in the lower left corner (the larger one is the SMA beam and the smaller one is the ALMA beam).}
    \label{fig:l314_mom1}
\end{figure}

In Fig.~\ref{fig:l314_mom1} we present the first-moment map in the 2.8~\kms\ component, obtained after   the  subtraction of the ``main'' Gaussian component observed in all other lines. This map is overlaid with contours of the 0.9~mm continuum emission measured with ALMA \citep{Zinchenko_et_al_2017_AA_606,Liu2018,Liu2020}. It is  
 important  to note that the angular resolution of the ALMA observations was several times higher than in the SMA observations. The plot shows a clear velocity gradient   the  similar to that observed in the $14_1 - 14_0$ A$^{-+}$ maser line in 2016 \citep{Zinchenko_et_al_2017_AA_606} and in other lines in this object \citep{Liu2020}.  Therefore, the probable maser emission arises in an extended region in the rotating disk-like structure around the protostar. 
 
\subsection{Methanol line emission observed with IRAM 30m}

According to Splatalogue\footnote{\url{https://splatalogue.online/advanced1.php}}
there are about 80 methanol transitions with the excitation energy of the upper level $< 1000$~K and the Einstein coefficient $A_E > 10^{-9}$ s$^{-1}$ in the observed frequency bands.  Three lines from  the $J_1 - J_0$ A$^{-+}$ series   with $J=11,13$  and  14 
are among them. One of the transitions is predicted as 
Class II methanol maser (at 330.793 GHz, \citet{Cragg_et_al_2005_MNRAS_360}), and a number of transitions are torsional exited ($v_t = 1,2$).

 No  more than 20 methanol transitions were detected even toward the most line-rich sources such as G049.489$-0.387$,  G111.542$+0.776$, and G192.600$-0.048$  with a confident signal to noise ratio $S/N\ga 5$ (Table~\ref{tab:transitions_list}). 
Methanol transitions that are blended with lines of others molecules were excluded from this list, since we could not conclude anything about 
the presence of their emission.

\begin{table*}
    \centering
    \caption{Methanol transitions that were observed with IRAM 30m. These are transitions 
 with their level energy below 1000 K and their Einstein coefficient $A_E > 10^{-9}$~s$^{-1}$
 (according to Splatalogue) and that have been detected toward  at least one source  with our sample. The numbers from 1 to 15 correspond to the source numbers in Table~\ref{tab:sources_list}.
 The "+" sign means a confident line detection ($\gtrsim 5\sigma$) and "?" means a tentative detection ($\gtrsim  2\sigma$). 
  The one transitions predicted as a Class II maser transitions by \citet{Cragg_et_al_2005_MNRAS_360} is marked with "MMII$^1$" in the "comment" column. The maser transitions newly identified by \citet{2019ApJ...881L..39B} are marked with "MMII$^2$".
 Lines from the $J_1 - J_0$ A$^{-+}$ series  are marked with "MMII??", since one of them was qualitatively classified as Class II maser \citep{Zinchenko_et_al_2017_AA_606}.
    }
    \label{tab:transitions_list}
\begin{tabular}{cccccccccccccccccccc}
\hline\noalign{\smallskip}
Frequency	&  $  E_{u}$&$A_{ij}$ & Notation &1&2&3&4&5&6&7&8&9&10&11&12&13&14&15&Comm. \\
  GHz 	&  K &s$^{-1}$&		                  \multicolumn{15}{c}{ } \\
\hline\noalign{\smallskip}                    
326.961232& 133.1& 1.29E-4&	$10_{-1}-9_0\, v_t0\, E$   & & & & & & & & &+&+&+&+&+&?&+& \\
327.317253& 218.7& 5.88E-5&	$12_2-11_3\, v_t0\, A^{--}$& & & & & & & &?&+&?&?&+&+& & &	\\	
327.407873& 492.8& 5.63E-5&	$17_5-18_4\, v_t0\, A^{++}$& & & & & & & & &?&?& & & & & &	\\	
327.440644& 492.8& 5.63E-5&	$17_5-18_4\, v_t0\, A^{--}$& & & & & & & &?&?& &?& & & & &	\\	
327.486835& 307.2& 5.62E-5&	$13_4-14_3\, v_t0\, E$     & & & & &?&?& &?&+&?&?&+&+& &+& \\	
329.632881& 218.8& 6.00E-5&	$12_2-11_3\, v_t0\, A^{++}$& &+&?&?&?&?& &?&+&+&+&+&+& &+&	\\	
330.172526& 810.7& 4.19E-5&	$11_3-12_4\, v_t0\, A^{--}$& & & & & & & & & & & &?& & & &  MMII$^2$ \\
330.172553& 810.7& 4.19E-5&	$11_3-12_4\, v_t0\, A^{++}$& & & & & & & & & & & &?& & & &   MMII$^2$ \\	
330.355512& 537.0& 6.42E-5&	$20_3-19_4\, v_t0\, A^{--}$& & & & & & & & & &?&?&?& & & &	\\	
330.793887& 146.3& 5.39E-5&	$8_{-3}-9_{-2}\, v_t0\,  E$&+&+&+&+&+&?&?&+&+&+&+&+&+& &+&   MMII$^1$  	\\	
331.220371& 320.6& 5.24E-5&	$16_{-1}-15_{-2}\,v_t0\, E$&?&+&+&+&+&?& &?&+&+&+&+&+& &+& 	\\
331.502319& 169.0& 1.96E-4&	$11_1-11_0\, v_t0\, A^{-+}$&+&+&+&+&+&+&+&+&+&+&+&+&+&?&+&	MMII??  \\
331.755099& 823.9& 1.27E-4&	$15_{-5}-16_{-6}\,v_t1\, E$& & & &?&?& & &?&+& &?&?& & & &   MMII$^2$ \\
332.996563& 614.5& 6.33E-5&	$22_{-2}-21_{-3}\,v_t0\, E$& & &+&?&?& & &?&+&?&?&?&+& &?&	\\	
333.864722& 125.5& 8.04E-7&	$9_{-1}-8_{2} \, v_t0\, E$ & & &?&?& &?& & &+&?& & &?& & & \\
333.864722& 125.5& 8.04E-7&	$9_1-8_{-2}	  \, v_t0\, E$ & & &?&?& &?& & &+&?& & &?& & & \\
334.426571& 314.5& 5.55E-5&	$3_0-2_1	  \, v_t1\, E$ &?&+&+&+&?&?& &?&+&?&+&+&+& &+&	\\	
342.729796& 227.5& 2.12E-4&	$13_1-13_0\, v_t0 \,A^{-+}$&?&+&+&+&+&+&?&+&+&+&+&+&+&?&+& 	MMII??  \\
343.599019& 624.0& 3.58E-5&	$13_{-1}-14_{-2}\,v_t1\, E$& & &?& & & & & &?& & &?& & &?&	  MMII$^2$ \\
344.109039& 419.4& 6.81E-5&	$18_2-17_3	\, 	  v_t0\, E$& &+&+&+&+&?& &?&+&+&+&+&+& &+&	\\
344.443433& 451.2& 9.35E-5&	$19_1-18_2\, v_t0 \,A^{++}$&?&?&+&+&?&?& &?&+&?&+&+&+& &+&	\\
344.970808& 761.6& 8.95E-5&	$12_7-11_6\, v_t1\, E$     & & & & & & & & &?& &?& & & &+& 	\\
345.903916& 332.7& 9.03E-5&	$16_1-15_2\, v_t0\, A^{--}$&?&+&+&+&+& & &+&+&+&+&+&+&?&+&		\\
345.919260& 459.4& 7.30E-5&	$18_{-3}-17_{-4}\, v_t0\,E$& &+&+&+&?& & &?&+&?&+&+&+&?&+&		\\
346.202719& 115.2& 2.18E-5&	$5_4-6_3\, v_t0 \,A^{--}$  &?&+&+&+&+&?&?&+&+&+&+&+&+&?&+&	 	\\
346.204271& 115.2& 2.18E-5&	$5_4-6_3\, v_t0 \,A^{++}$  &?&+&+&+&+&?&?&+&+&+&+&+&+&?&+&	 \\
349.106997& 260.2& 2.20E-4&	$14_1-14_0\, v_t0\, A^{-+}$&?&+&+&+&+&+&?&+&+&+&+&+&+&?&+&	MMII?? \\
\hline
\end{tabular}
\end{table*}

It is worth noting that only the lines of  the $J_1 - J_0$ A$^{-+}$ series and the line blend $5_4 - 6_3$ A$^{--,++}$ are detected in all sources.   The brightest non-blending lines that are visible in 13 out of the 15 sources are presented in Fig.~\ref{fig:all_sou_sp}.

 \begin{figure*}
    \includegraphics[width=0.90\linewidth]{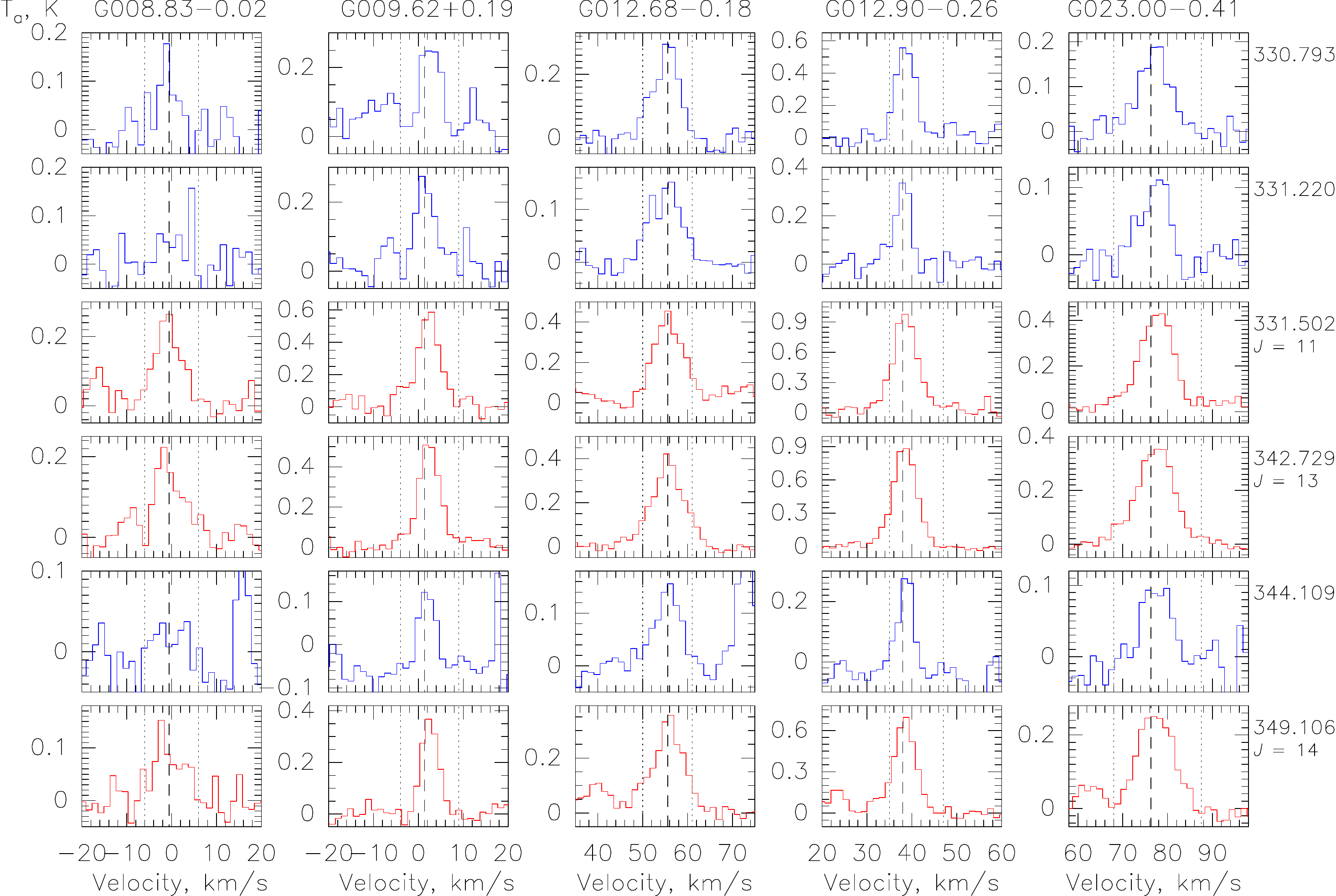}
	\includegraphics[width=0.90\linewidth]{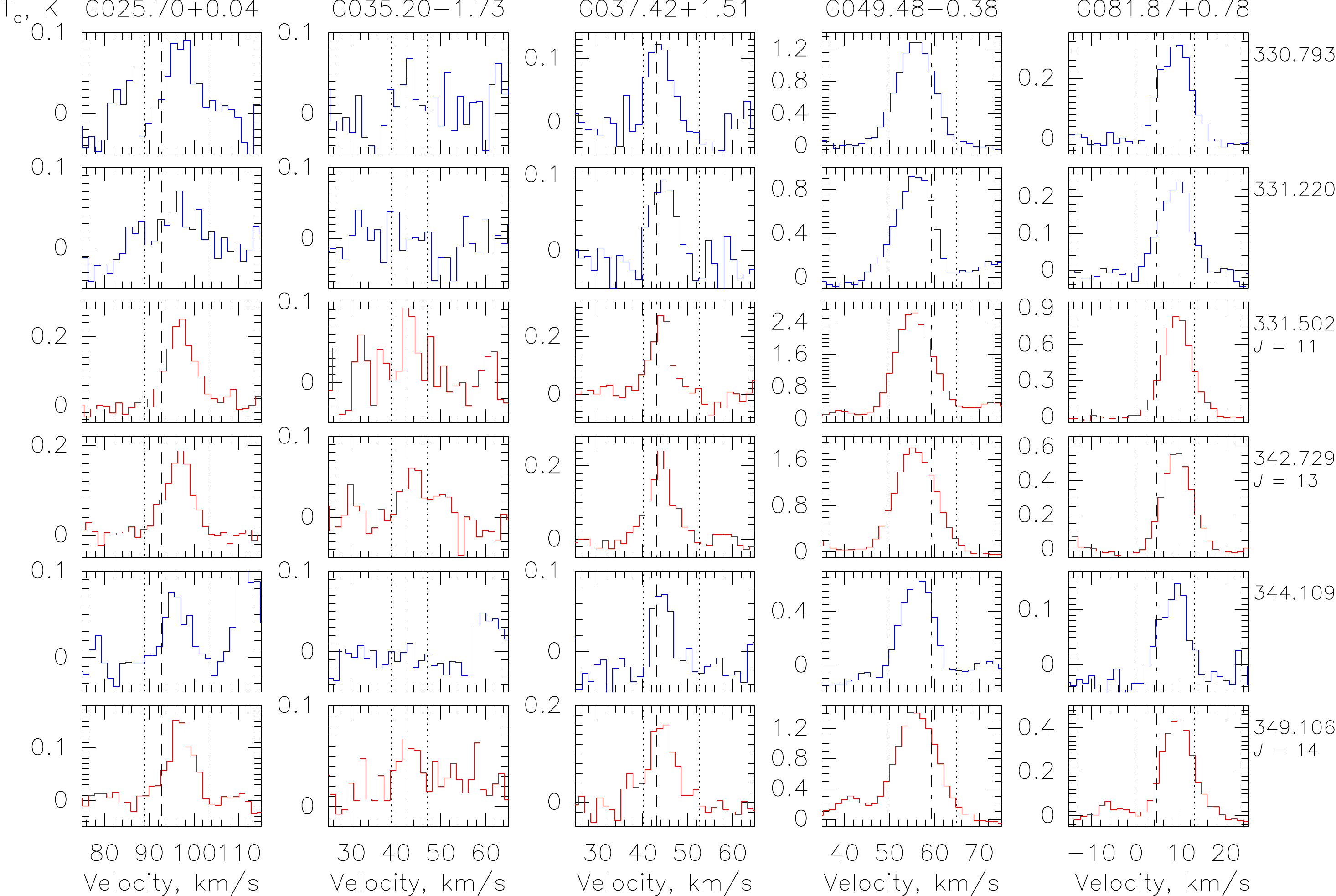}
    \caption{CH$_3$OH line spectra  obtained from the IRAM 30m observations . The lines of the $J_1 - J_0$ A$^{-+}$ series are plotted in red  while  other lines in blue. 
      The velocity range of the 6.7~GHz maser emission and its peak velocity  (see Table \ref{tab:sources_list})   are labeled  with dotted and dashed lines,   respectively .
    Peak velocities of the sources observed with RT32m are plotted in dotted-dashed lines.
    The line frequencies in GHz are indicated on the right side. The quantum numbers $J$ for the transitions from the $J_1 - J_0$ A$^{-+}$ series are shown, too.
}  \label{fig:all_sou_sp}
\end{figure*}
\addtocounter{figure}{-1}
 \begin{figure*}
	\includegraphics[width=0.90\linewidth]{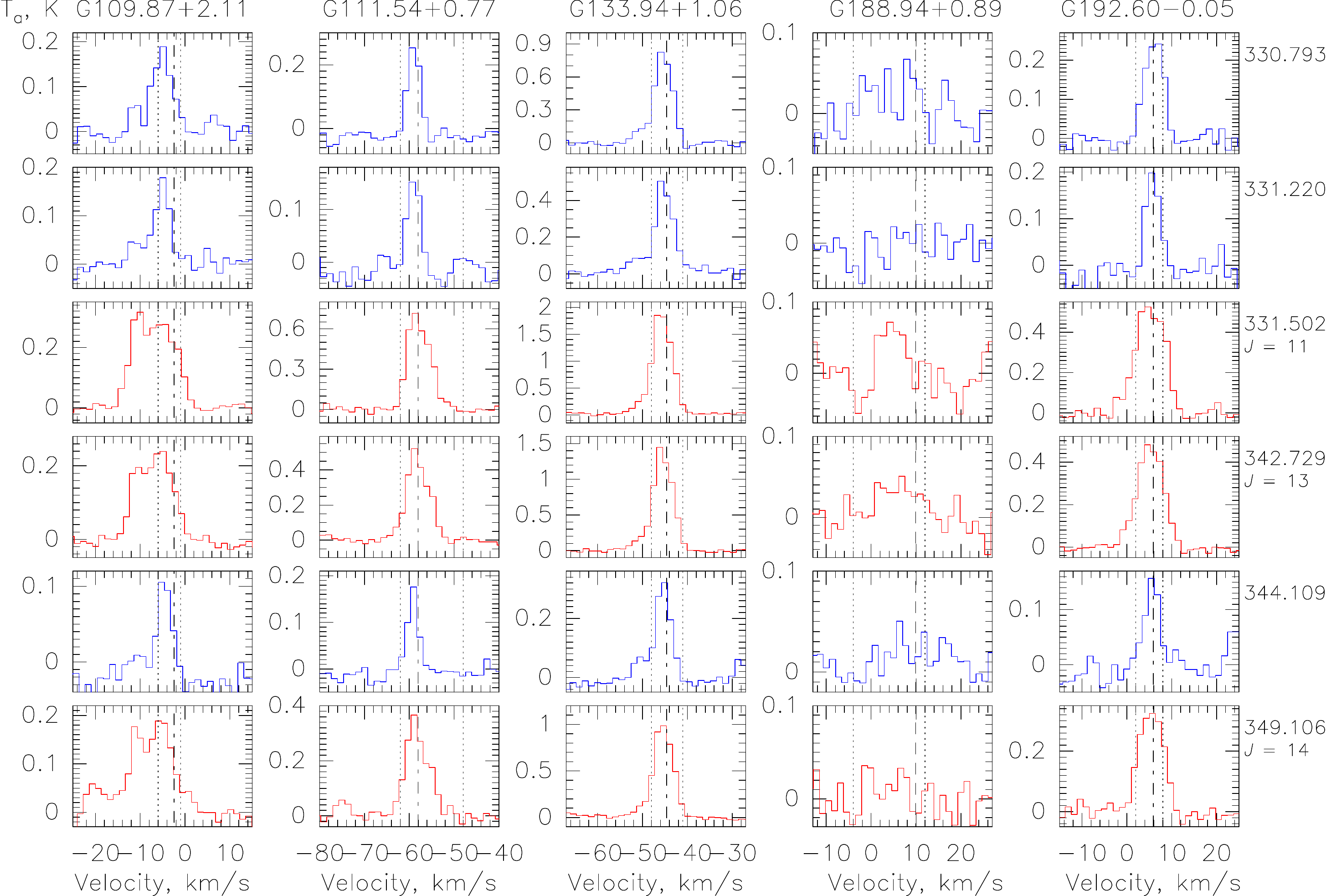}
    \caption{(Continued.) }  \label{fig:all_sou_sp_cont}
\end{figure*}

\begin{figure}
    \includegraphics[width=0.90\columnwidth]{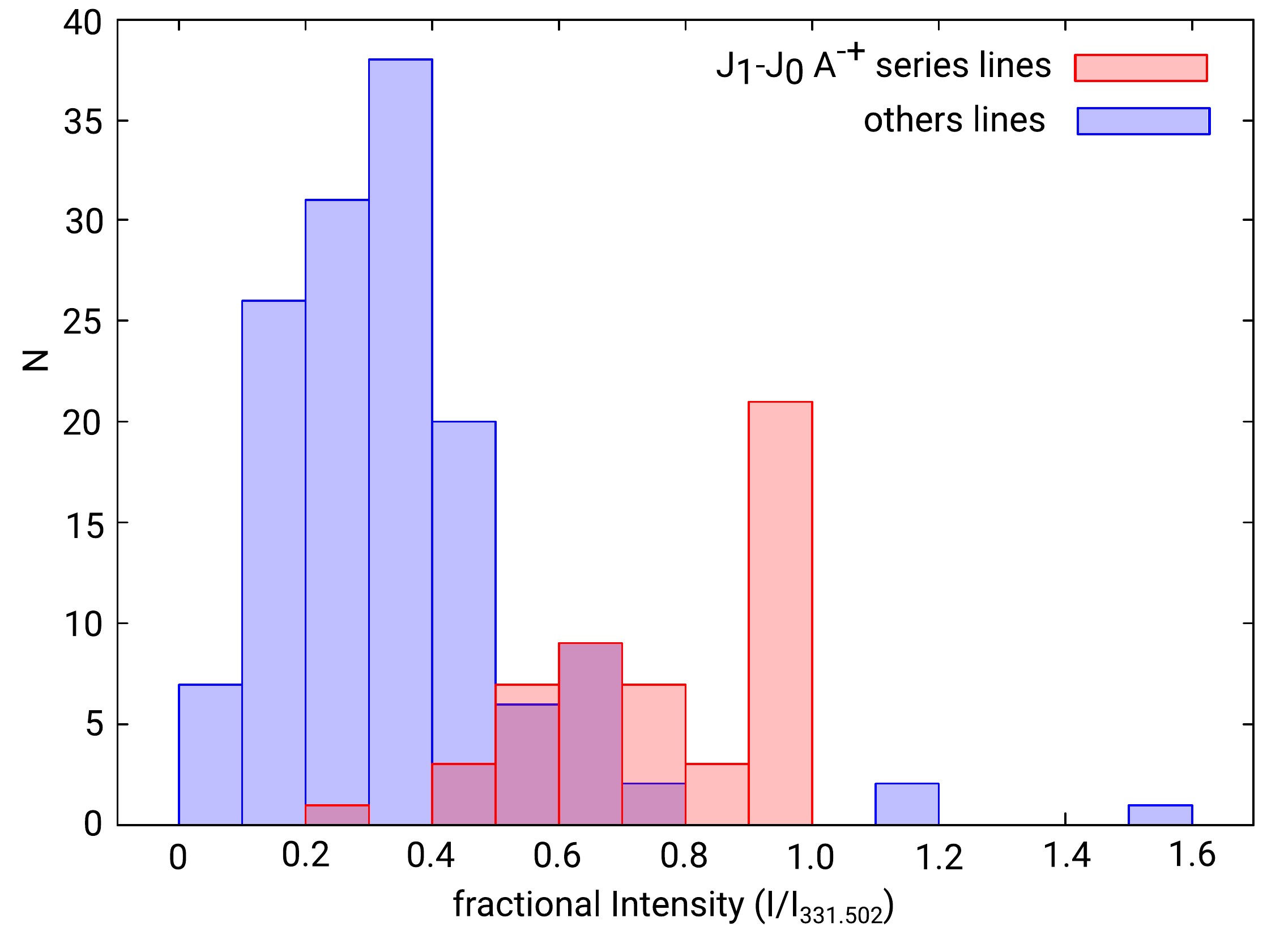}
    \includegraphics[width=0.90\columnwidth]{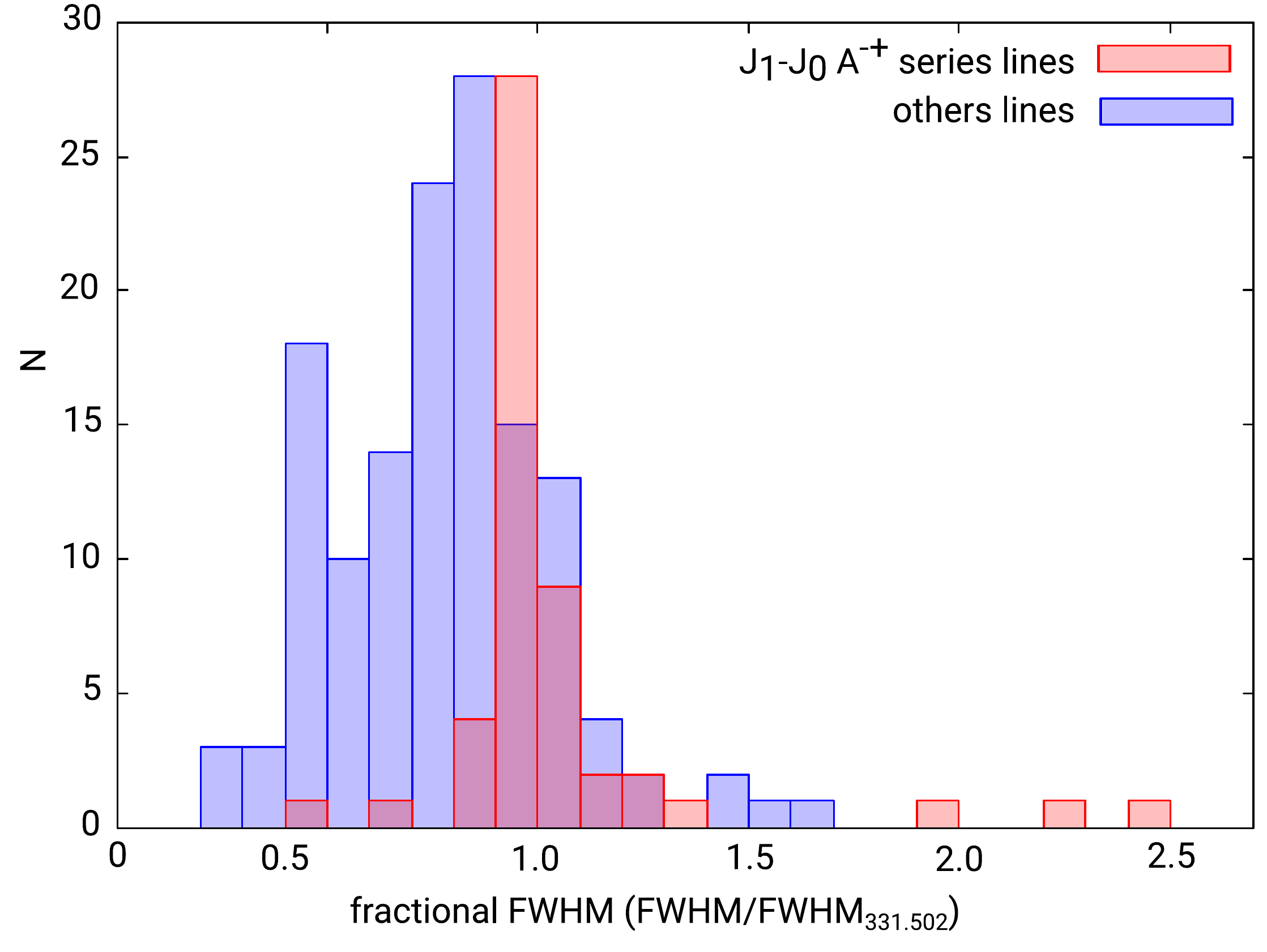}
    \caption{Top panel: The distribution of the lines intensities relative to the intensity of the brightest transition in the $J_1 - J_0$ A$^{-+}$ series at 331.502 GHz with $J=11$. Lower panel: The distribution of the fractional line widths in FWHM relative to that of the same ($J=11$) transition.
    }  
    \label{fig:histograms}
\end{figure}

As can be seen in Fig.~\ref{fig:all_sou_sp}, toward most of the sources observed at 0.8 mm the methanol line emission appears   within  
the same velocity intervals as   the  Class II methanol maser at 6.7 GHz. However, no obvious bright maser effect is registered in any source in the sample. 
All observed methanol lines are rather broad (with their line widths ranging from 3 to 10~\kms). 
It is noteworthy that the lines of the $J_1 - J_0$ A$^{-+}$ series are the brightest lines everywhere. 
 This is most clearly seen in Fig.~\ref{fig:histograms} (upper histogram), which shows the ratio of the intensities of all observed lines in the each of considered sources to the brightest line from the series  $J_1 - J_0$ A$^{-+}$ with $J=11$ at 331.502~GHz. 
Only the line at 326.961 GHz ($10_{-1}-9_0\, E$) in the  sources G049.489$-0.387$, G081.871$+0.780$, and G109.870$+2.114$ has higher intensities. This transition belongs to the  $J_{-1}-J-1_0\, E$ series and is considered as a Class I methanol maser candidate \citep{2012IAUS..287..433V}. In the sources G111.542$+0.776$ and G133.947$+1.064$ the intensities of this line are comparable with the intensities of the $11_1 - 11_0$ A$^{-+}$ line and for other sources we cannot say anything about the intensity of the $10_{-1}-9_0\, E$ line due to large noise in the corresponding part of the band.  
  We also note  that we do not include in the histograms the lines at 346.202 and 346.204 GHz ($5_4-6_3 A^{--}$ and $5_4-6_3 A^{++}$), since they are blended in all sources and we could not  separate  them correctly. As it can be seen in   the lower histogram  of Fig.~\ref{fig:histograms}, 
   where the FWHM relations are presented, 
the lines of the series are broader then other methanol lines for all sources.  

In some sources one can see additional spectral components out of the range of the 6.7~GHz emission. 
For instance, in the source G133.947+1.064  an extra emission  at $\sim -50$~\kms. 
The most  notable spectral component in some methanol line emission out of 6.7~GHz emission ranges is seen in the G109.870+2.114 (Cep~A). 
A significant differences between the line profiles of the $J_1 - J_0$ A$^{-+}$ series and other methanol lines are detected in the source  G111.542+0.776 (NGC 7538C) spectra. 

In G192.600-0.048 an enhanced emission at velocity about 3~\kms\ in the $11_1 - 11_0$ A$^{-+}$ transition at 331.5~GHz can be seen. This emission is marginally noticeable in the spectrum of the $13_1 - 13_0$ A$^{-+}$   line    
(Figs.~\ref{fig:all_sou_sp}). 

Under the  assumption of low optical depth in the lines and LTE conditions, we can plot rotational diagrams for all sources. 
The standard equation for the rotational diagram method (i.e. \citep{1991ApJS...76..617T})  was used:
$$
\ln\frac{3kW}{8\pi^3\nu S\mu^2}=\ln\frac{N}{Q}-\frac{E_u/k}{T_{rot}},
$$
where $k$ is the Boltzmann constant,    
$W$ is the observed integrated line intensity,   
$\nu$ is the frequency of the transition,  
and $S\mu^2$ is the product of the  
line strength and the square of the electric dipole moment,    
$E_u/k$ is the upper level energy in temperature scale,   
$Q$ is the partition function at temperature $T_{rot}$ that was computed as $Q = 0.38T_{rot}^{1.76}$ (according to the fitting of the data from the JPL database\footnote{\url{https://spec.jpl.nasa.gov}} ). 

This method can be applied only for the sets   of  transitions with thermal excitation. Thus,   the  transitions suspected in anomalous excitation were excluded from the consideration.  They are the transitions from the $J_1 - J_0$ A$^{-+}$ series,   the  transition at 330.793 GHz, which are assumed to be Class II methanol maser,   the transition at 326.961 GHz, 
which are assumed to be Class I methanol maser, and the transition at 334.426 GHz,   
which is   a  torsional excited transition. Blended lines at 346.202 and 346.204 GHz were excluded too. Moreover,  only those  lines with S/N ratio greater than 3 were  considered.
 
\begin{figure*}
    \includegraphics[width=0.85\linewidth]{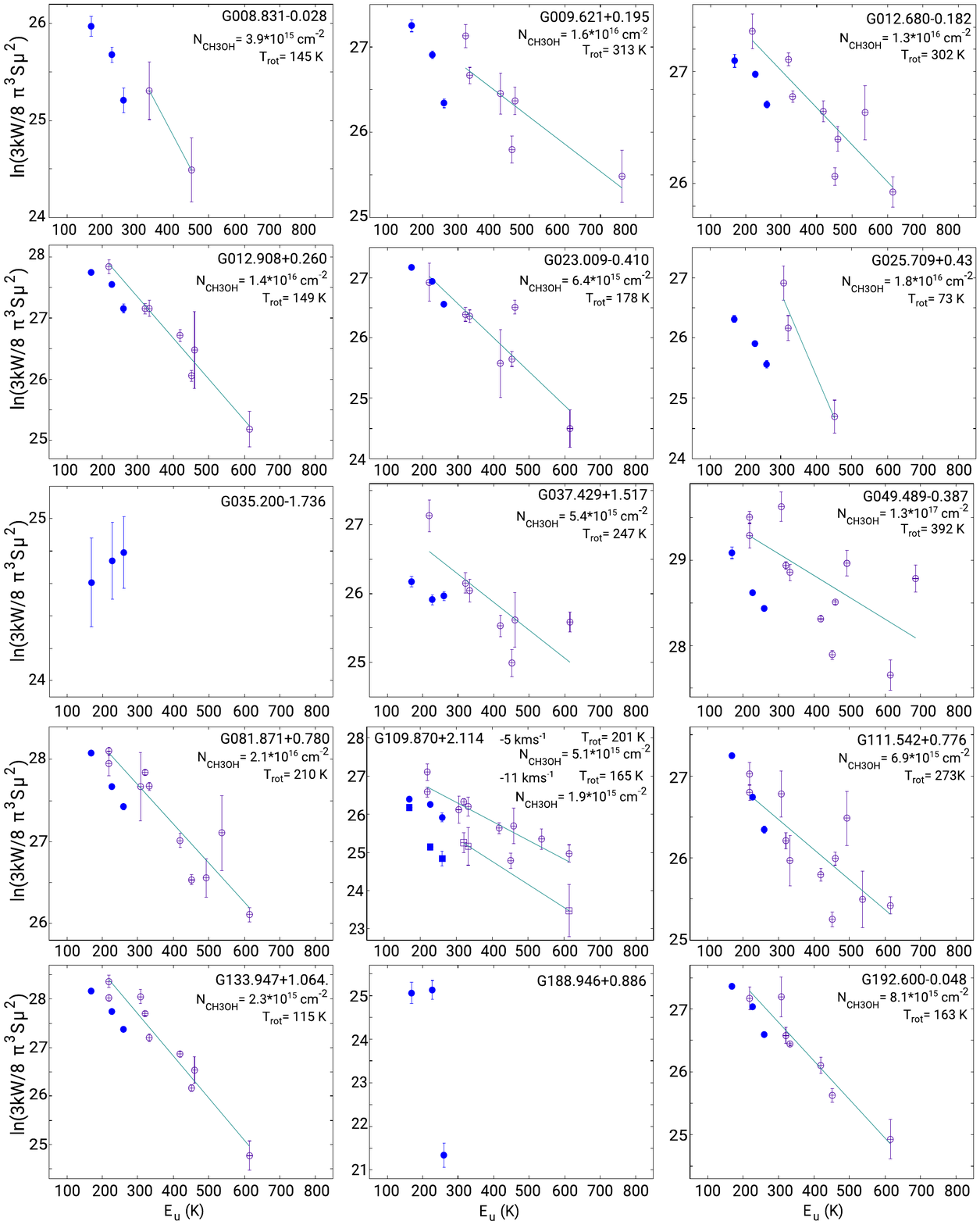}
    \caption{Rotational diagrams for transitions with confident (S/N > 3) detection in 1~Gaussian fitting from the IRAM 30m observations. Data points from one-Gaussian fitting for all transitions are marked by open circles except for G109.870+2.114, in which the two-Gaussian fitting was used and the additional component is marked by open squares. The $J_1 - J_0$ A$^{-+}$ lines, which were not use in the rotational diagram analysis, are marked in solid blue. 
}  \label{fig:rot_dia}
\end{figure*}

According to the  rotational diagrams  shown in  Fig.~\ref{fig:rot_dia}, we found rotational temperatures,  $T_{rot}$,  ranging from $\sim$70~K to $\sim$390~K, and methanol column densities, $N_{CH_3OH}$, ranging from $2\cdot10^{15}$~cm$^{-2}$ to $10^{17}$~cm$^{-2}$ for the sources under consideration.

\subsection{6.7 GHz masers observed with RT-32}

The maser emission at 6.7 GHz observed  with RT-32, Irbene, for all 6 sources has about the same velocity intervals that were presented in \citet{Hu_et_al_2016}.  
Some differences in the velocities of the main components   
may be related to the masers variability. 
Remarkably, for all 6 sources observed simultaneously both   with  RT-32 and  with IRAM 30m the velocity of the brightest  component at 6.7 GHz does not coincide exactly   
with the peak velocity of other methanol lines.    

\begin{figure}
	\includegraphics[width=0.99\linewidth]{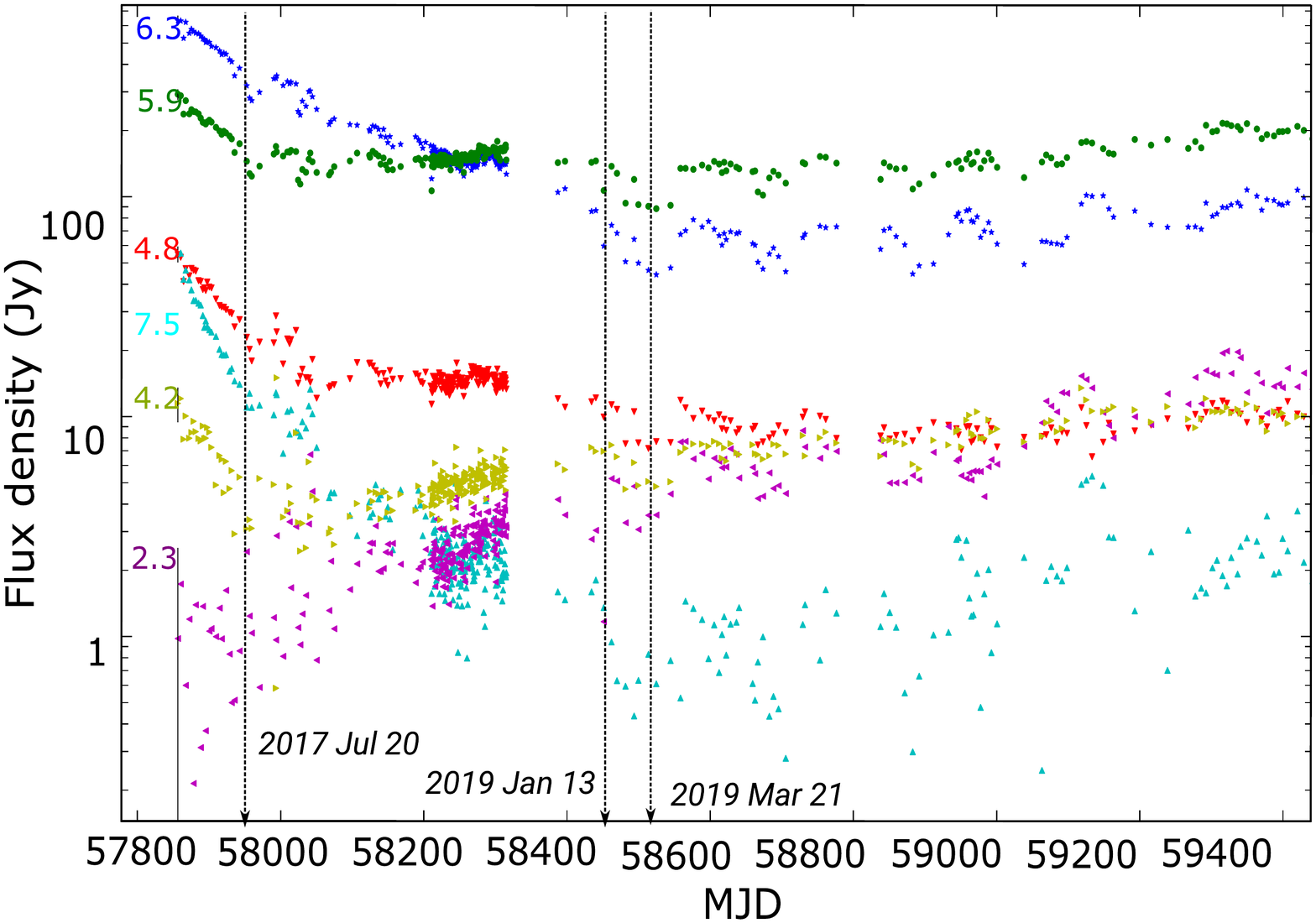}
    \caption{ Flux densities of the 6.7 GHz maser in S255IR obtained from the monitoring program with RT-32, Irbene. The maser components at different velocities (labeled on the left in \kms) are plotted in different colors. The epochs of observations discussed in this paper are indicated with arrows.   Calibration uncertainties are shown as the black vertical lines in the beginning of the time scale. }  \label{fig:monitoring}
\end{figure}

We note that the observations on 09-10 January 2019 are a part of the long-term monitoring program of these sources with RT-32. Based on those results we can conclude that 
about two years before   and about two years later   the January 2019 observations, 
no strong variations of the 6.7 GHz maser emission were recorded (see an example in Fig. \ref{fig:monitoring}).

\section{Modeling methanol emission  
} \label{sec:model}
\subsection{Methods of analysis}

To interpret the observed brightness temperatures of the methanol lines, we resort to the database compiled by \citet{2018msa..conf..276S}, which tabulates the population numbers for quantum energy levels of methanol of ground and torsionally excited levels with $v_t$ up to 2. 
To evaluate the   population numbers  for the quantum energy levels the large velocity gradient (LVG) approach was employed over a 5-dimension grid of physical  parameters, which include the gas kinetic temperature ($T_k$, K), the hydrogen number density ($n_{\rm H_2}$, cm$^{-3}$), the methanol specific column density ($N_{\rm CH_3OH}/\Delta V$,~cm$^{-3}$s), the methanol relative abundance ($N_{\rm CH_3OH}/N_{\rm H_2}$) and the line width ($\Delta V$~\kms). 
Dust emission and absorption within the emission region  were taken into account in the way described  in \citet{2004ApJ...609..231S}. It  is assumed that the dust particles are intermixed with gas homogeneously and have the same physical temperature. 
Details of the LVG calculation can be found in \citet{2015ApJ...810...10Z, 2021MNRAS.503..633K}. 

Parameters in the database are varied in the ranges from 10 to 600 K for $T_k$, from 3.0 to 9.0 for $\lg(n_{\rm H_2})$, from 7.5 to 14.0 for $\lg(N_{\rm CH_3OH}/\Delta V)$,   from $-9.0$ to $-5.5$ for $\lg(N_{\rm CH_3OH}/N_{\rm H_2})$, and finally at 1, 3, 5~\kms for $\Delta V$.
A beam filling factor 
for the methanol emission ($f = 10-100\%$)
was  included  as an additional parameter in the methanol line intensity analysis. Since the line width is about 3~\kms in the sources under consideration, this value was fixed for all models here. 
 
To extend the analysis of the methanol emission and to obtain the confidence intervals, we applied Bayesian approach \citep[e.g.][]{Ward2003}. Methanol model intensities ($T_{{\rm m}\,i}(p)$) for regular network of parameters:
$$ p = \left(T_{k} , N_{\rm CH_3OH}/\Delta V, n_{\rm H_2}, N_{\rm CH_3OH}/N_{\rm H_2}, f\right)$$ were calculated using the database of the population numbers. 
 
The probability to observe $N$ methanol lines with intensities $T_{o\,i}$ and uncertainties $\sigma_i$ using $p$ set of parameters can be calculated as:
$$
P(T_{\rm o}|p) = \prod_i^N{\frac{1}{\sqrt{2\pi}\sigma_i}\,e^{-\frac{1}{2}\left(\frac{T_{{\rm o}\,i} - T_{{\rm m}\,i}(p)}{\sigma_i}\right)^2}}.
$$
Integrating over each of the parameters, we obtained  the Bayesian probability function and estimate the confidence intervals.

\subsection{Emission in the $J_1 - J_0$ A$^{-+}$ line series observed with the SMA in the S255IR}

As mentioned above,  6 lines from the $J_1 - J_0$ A$^{-+}$ series were observed with the SMA in S255IR. 
  We attempted to find   the  physical conditions under which all these lines can be fitted sufficiently well   not more than $3\sigma$~rms  by a single component but could not find any set of parameters which provides such a fit. 
Intensities of the lines with $J=6$ and 7 are underestimated in the models while those of the   $J=4,5, 13, 14$ lines have been fitted  reasonably well.  

\begin{figure}
	\includegraphics[width=1\columnwidth]{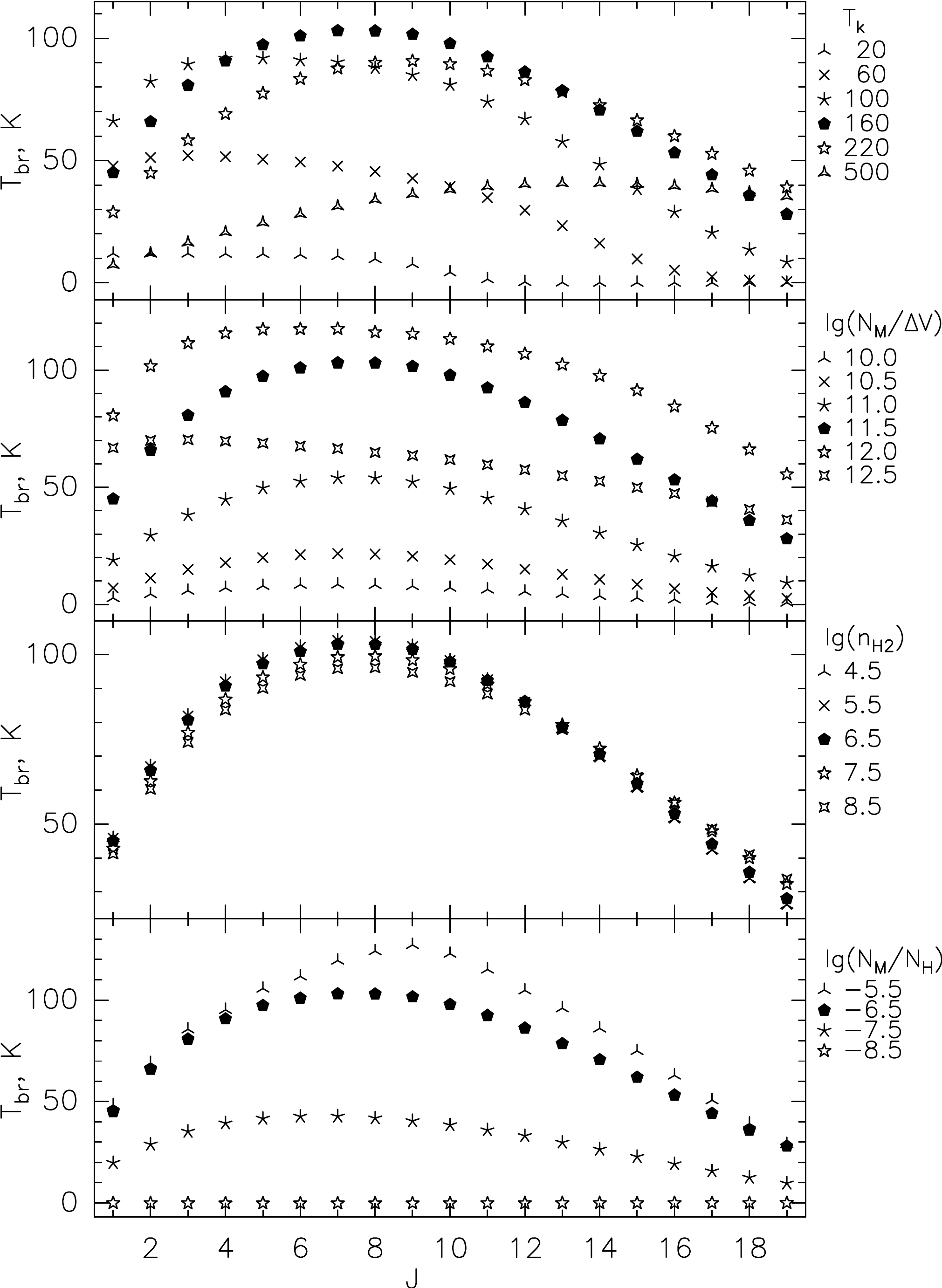}
    \caption{Dependence of brightness temperatures $T_{br}$ on the kinetic temperature, molecular hydrogen number density, methanol specific column density, and fractional abundance 
    for the $J_1 - J_0$ A$^{-+}$ series of transitions. }  \label{fig:J1_J0Amp_runs}
\end{figure}

 To find the conditions under which the lines with $J=7$ is the brightest in the series, we study dependence of the modeled  peak brightness of the lines of the series on physical parameters.  
We used a fiducial model with parameters $T_k = 160$~K, $\lg (N_{CH_3OH}/\Delta V) = 11.5$, $\lg (n_{H_2} = 6.5$, $N_{CH_3OH}/N_{H_2} = 10^{-6.5}, \Delta V = 3$~\kms, $f=100\%$. These values are consistent with the best-fit model for a set  where transition  with $J=7$ was excluded from analyses.
Varying parameters of the fiducial model one at a time we investigated dependence of the brightness temperature ($T_{br}$) of the lines on the parameters of the model.  
It can be seen from Fig. \ref{fig:J1_J0Amp_runs} that, with increasing temperature, the brightness maximum shifts towards transitions with higher $J$ numbers.  

Some dependence of the $J$ number of the brightest line on the
methanol specific column density and fractional abundance does exist.
We have found that the $J$ number of the brightest line does not change with variation of the number density in our models. 
This indicates that radiative processes play a major role in the excitation of the transitions of the $J_1 - J_0$ A$^{-+}$ series. 
This resembles  Class II methanol maser excitation, which is predominantly radiative.

\subsection{Conditions for maser emission in the $J_1 - J_0$ A$^{-+}$ line series}\label{sec:maser_conditions_J1_J0Amp}

Since we found in the spectral profile of the $7_1 - 7_0 A^{-+}$ line at 314.860 GHz a  feature that is similar to the one 
detected by \citet{Zinchenko_et_al_2017_AA_606} and we could not explain it in the simple LVG model without beaming and other properties characteristic of the bright masers, we confined ourselves to qualitative exploration of possibilities for the maser amplification in the transitions of the $J_1 - J_0$ A$^{-+}$ series.
Consequently, we searched for conditions at which an inversion in the   population numbers of transitions  occurs.

There is a number of models where the level populations  of the $J_1 - J_0$ A$^{-+}$ series transitions are inverted in our calculations. Using the database \citep{2018msa..conf..276S} we found out models with $\tau<-0.01$  and $<-0.1$ (where maser amplification exceeds 1\% and 10\% for transitions with the highest inversion in the series) for 19 transitions of the series (Table \ref{tab:J1_J0Amp_mod_res}). 

Transitions of the  $J_1 - J_0$ A$^{-+}$ series could be inverted with gas kinetic temperatures from 110 K up to 600 K (the upper edge of the gas kinetic temperatures in our database). 
Methanol specific column density varies from  $2.5\cdot10^{8}$ to $5 \cdot 10^{11}$ cm$^{-3}$s. The methanol abundances should exceed  $10^{-7}$.  Remarkably,  there are no constraints on the hydrogen number density within the range of the database. 
This is consistent with the classification of the masers in this series as Class~II \citep[as suggested in][]{Zinchenko_et_al_2017_AA_606}, which are characterized by radiative excitation.

 According to \citet{Cragg_et_al_2005_MNRAS_360} different maser lines series have different sensitivity to physical parameters. In our model for  the Class II maser at 6.7 GHz an inversion with $\tau < -1 $ occurs with gas kinetic temperatures from 40~K up to 90~K, methanol specific column density $>10^{11}$~cm$^{-3}$s, methanol abundances $\ge 3\cdot10^{-7}$ and hydrogen number density $<4\cdot10^{6}$~cm$^{-3}$.

Within the scope of this paper we provide only qualitative explanation of the maser occurrence in the transitions of the $J_1 - J_0$ A$^{-+}$ series. However, we can mention directions in which the quantitative explanation of this maser phenomenon can be sought for. These are, first of all, computations within much more elaborated model which takes into account the differences in the temperatures of the pumping emission and the gas temperature within the maser sources,  both of which are very important for the pumping of the Class~II methanol masers \citep{1994A&A...291..569S, Cragg_et_al_2005_MNRAS_360}.
Another point which has to be taken into account for explaining the maser intensities is the maser beaming  effect which greatly affects methanol maser intensities and their ratios \citep{2018IAUS..336...57S}.

However, we have to mention this can be not enough because available models did not show considerable maser emission in the $J_1 - J_0$ A$^{-+}$ series. This is probably related to insufficient number of the methanol energy levels in the model computations which is very important for the maser modelling or some simplifications in the available modelling  \citep{2012IAUS..287...13S}.

Taking into account this high complexity of the quantitative modelling of the masers in the $J_1 - J_0$ A$^{-+}$ series of transitions, in the current paper we are sticked to the qualitative analysis of this maser phenomenon.
 
\begin{table*}
	\centering
\caption{  Physical condition extracted from the database compiled by \citet{2018msa..conf..276S}, 
under which inverted levels population could happen (with $\tau < -0.01 $ and $\tau < -0.10 $). 
The line width was fixed at 3~\kms.  Parameter boundaries under consideration are highlighted in the second row of the header. 
}  \label{tab:J1_J0Amp_mod_res}
\begin{tabular}{c c c c c c c c c c c}
\hline 
 Frequency, & $J$&$E_u$ &\multicolumn{2}{c}{$T_k$, K}  &\multicolumn{2}{c}{$\lg(N_{\rm CH_3OH}/\Delta V$, cm$^{-3}$\,s)}&\multicolumn{2}{c}{$\lg(n_{\rm H_2}$, cm$^{-3})$}&\multicolumn{2}{c}{$\lg(N_{\rm CH_3OH}/N_{\rm H_2})$}\\ 
    GHz   &    &  K            &\multicolumn{2}{c}{$10-600$}  &\multicolumn{2}{c}{$7.5 - 14.0$}&\multicolumn{2}{c}{$3.0 - 9.0$}&\multicolumn{2}{c}{$-9.0- -5.5$}\\ 
         &      &            &$\tau < -0.01$&$\tau < -0.10$ &$\tau < -0.01$& $\tau < -0.10 $ &$\tau<-0.01$&$\tau < -0.10$      &$\tau < -0.01$& $\tau < -0.10 $  \\
\hline
 303.367 & $1$&16.9 & $110 - 600$ &	  $140-370$ &$8.5 - 11.2$  & $9.7- 10.8$ &$3.0 - 8.50$& $3.0 - 6.50$ & $-6.5 - -5.5$&  $-5.5$\\
 304.208 & $2$&21.6 & $110 - 600$ &   $140-520$ &$8.4 - 11.2$  & $9.4 -10.9$ &$3.0 - 8.75$& $3.0 - 7.00$ & $-7.0 - -5.5$&  $-5.5$\\      
 305.473 & $3$&28.6 & $120 - 600$ &   $140-600$ &$8.4 - 11.3$  & $9.3- 11.0$ &$3.0 - 8.75$& $3.0 - 7.50$ & $-7.0 - -5.5$&  $-6.0 - -5.5$\\
 307.166 & $4$&38.0 & $120 - 600$ &   $140-600$ &$8.3 - 11.3$  & $9.3- 11.0$ &$3.0 - 9.00$& $3.0 - 7.75$ & $-7.0 - -5.5$&  $-6.0 - -5.5$\\  
 309.290 & $5$&49.7 & $130 - 600$ &   $140-600$ &$8.3 - 11.3$  & $9.3- 11.1$ &$3.0 - 9.00$& $3.0 - 7.75$ & $-7.0 - -5.5$&  $-6.0 - -5.5$\\  
 311.853 & $6$&63.7 & $140 - 600$ &   $150-600$ &$8.4 - 11.3$  & $9.4- 11.1$ &$3.0 - 9.00$& $3.0 - 8.00$ & $-7.0 - -5.5$&  $-6.0 - -5.5$\\   
 314.860 & $7$&80.1 & $140 - 600$ &   $150-600$ &$8.5 - 11.4$  & $9.5- 11.1$ &$3.0 - 9.00$& $3.0 - 8.00$ & $-7.0 - -5.5$&  $-6.0 - -5.5$\\  
 318.319 & $8$&98.8 & $150 - 600$ &   $160-600$ &$8.6 - 11.4$  & $9.6- 11.1$ &$3.0 - 9.00$& $3.0 - 8.00$ & $-7.0 - -5.5$&  $-6.0 - -5.5$\\   
 322.239 & $9$&119.9 & $150 - 600$ &   $170-600$ &$8.7 - 11.4$  & $9.7- 11.2$ &$3.0 - 9.00$& $3.0 - 8.25$ & $-7.0 - -5.5$&  $-6.0 - -5.5$\\   
 326.631 & $10$&143.3 & $160 - 600$ & $170-600$ &$8.7 - 11.4$  & $9.7- 11.2$ &$3.0 - 9.00$& $3.0 - 8.50$ & $-6.5 - -5.5$&  $-6.0 - -5.5$\\
 331.502 & $11$&169.0 & $160 - 600$ & $180-600$ &$8.8 - 11.4$  & $9.8- 11.1$ &$3.0 - 9.00$& $3.0 - 8.50$ & $-6.5 - -5.5$&  $-5.5$\\  
 336.865 & $12$&197.1 & $170 - 600$ & $200-600$ &$8.9 - 11.4$  & $9.9- 11.1$ &$3.0 - 9.00$& $3.0 - 8.50$ & $-6.5 - -5.5$&  $-5.5$\\  
 342.730 & $13$&227.5 & $170 - 600$ & $210-600$ &$8.9 - 11.4$  & $9.9- 11.1$ &$3.0 - 9.00$& $3.0 - 8.50$ & $-6.5 - -5.5$&  $-5.5$\\  
 349.107 & $14$&260.2 & $170 - 600$ & $220-600$ &$9.0 - 11.4$  & $10.0- 11.1$ &$3.0 - 9.00$& $3.0 - 8.50$ & $-6.5 - -5.5$&  $-5.5$\\  
 356.007 & $15$&295.3 & $180 - 600$ & $230-600$ &$9.0 - 11.4$  & $10.0- 11.1$ &$3.0 - 9.00$& $3.0 - 8.50$ & $-6.5 - -5.5$&  $-5.5$\\  
 363.440 & $16$&332.6 & $180 - 600$ & $250-600$ &$9.1 - 11.4$  & $10.0- 11.0$ &$3.0 - 9.00$& $3.0 - 8.50$ & $-6.5 - -5.5$&  $-5.5$\\   
 371.415 & $17$&372.4 & $190 - 600$ & $260-600$ &$9.1 - 11.4$  & $10.2- 11.0$ &$3.0 - 9.00$& $3.0 - 8.50$ & $-6.5 - -5.5$&  $-5.5$\\   
 379.940 & $18$&414.4 & $200 - 600$ & $290-600$ &$9.2 - 11.3$  & $10.3- 11.0$ &$3.0 - 9.00$& $6.5 - 8.25$ & $-6.5 - -5.5$&  $-5.5$\\  
 389.021 & $19$&458.7 & $210 - 600$ & $330-600$ &$9.2 - 11.3$  & $10.4- 10.9$ &$3.0 - 9.00$& $7.0 - 8.25$ & $-6.5 - -5.5$&  $-5.5$\\  
\hline
\end{tabular}
\end{table*}

\subsection{Estimation of the physical condition in the sample sources }

In the rotational diagrams (Fig.~\ref{fig:rot_dia}) one can see rather large deviations from LTE for most of the sources. 
Thus we use the LVG approximation for estimating the physical parameters of the sources under consideration. For most of the sources the values of $T_{obs}$ obtained from the single Gaussian fitting were used.
For G109.870+2.114 two Gaussian fitting was used since there are two clearly distinguished spectral components in the methanol spectra. Moreover, for G111.542+0.776 we computed the parameters both by one Gaussian fitting  and two Gaussian fitting as will be discussed in Section~\ref{sec:ouG111p54}. 
Using    the  methanol line intensities we estimated  the  physical parameters corresponding to the $\chi^2$ minimum and their 68\% confidence intervals.   We present the results in Tab.~\ref{tab:phis_params_res}. The same set of the 11 brightest transitions was used in the analysis for all sources apart from the components   G111.542+0.776 ($-56.4$ \kms) and G111.542+0.776 ($-59.0$ \kms). For them we used all 16 detected lines. 
The line width value of 3 \kms for all sources under consideration was fixed   
  in  all models. 
For all sources we made sure that the intensities of unregistered lines do not exceed $1\sigma$.

\begin{table*}
\centering
\caption{  Physical conditions inferred from the LVG analysis with the minimum $\chi^2$ within the parameter space under consideration listed in the last header row. The 68\% confidence intervals for the parameters are shown in brackets.  Parameter boundaries under consideration are highlighted in the second row of the header. }
\label{tab:phis_params_res}
\small
\begin{tabular}{lrrrrr}
\hline 
Source           &$T_k$, K   &lg($N_{\rm CH_3OH}/\Delta V$, cm$^{-3}$\,s) &lg($n_{H2}$, cm$^{-3}$)        &lg($N_{\rm CH_3OH}/N_{H_2}$)   &$f$ \% \\
                        & $10-600$ & $7.5-14.0$ &   $3.00-9.00$   &  $-9.0- -5.5$  & $10-100$ \\
\hline 
G008.831$-0.028$          &   $50(25-80)$        &$11.5(9.1-12.1)$    &$5.50(3.00-9.00)$   &$-8.0(-9.0 - -7.0)$  &$30(10-100)$\\
G009.621$+0.195$          &   $270(250-290)$     &$12.5(12.4-12.7)$   &$5.75(5.30-9.00)$   &$-7.0(-7.4 - -6.6)$  &$20(10-45)$\\
G012.680$-0.182$          &   $140(105-170)$     &$12.0(11.9-12.1)$   &$5.75(5.40-9.00)$   &$-7.5(-7.8 - -7.1)$  &$30(25-40)$\\
G012.908$-0.260$          &   $160(150-170)$     &$12.2(12.1-12.3)$   &$5.50(5.00-7.75)$   &$-7.0(-7.3 - -6.7)$  &$10(10-20)$\\
G023.009$-0.410$          &   $190(170-210)$     &$12.0(11.9-12.1)$   &$9.00(8.10-9.00)$   &$-7.5(-7.8 - -7.1)$  &$20(10-30)$\\
G025.709$+0.043$          &   $150(90-280)$      &$11.5(9.0-11.7)$    &$5.75(3.00-9.00)$   &$-8.0(-9.0 - -6.6)$  &$30(10-100)$\\
G035.200$-1.736$          &   $30(20-110)$       &$12.1(8.7-14.0)$    &$9.00(3.00-9.00)$   &$-9.0(-9.0 - -6.0)$  &$60(10-100)$\\
G037.429$+1.517$          &   $200(175-455)$     &$13.0(12.5-13.2)$   &$5.75(5.50-9.00)$   &$-6.5(-7.4 - -6.2)$  &$10(10-20)$\\
G049.489$-0.387$          &   $220(210-230)$     &$12.5(12.4-12.6)$   &$5.50(5.30-6.40)$   &$-7.0(-7.3 - -6.6)$  &$80(70-90)$\\
G081.871$+0.780$          &   $180(155-190)$     &$12.1(11.9-12.2)$   &$9.00(6.50-9.00)$   &$-7.5(-7.8 - -7.2)$  &$80(30-95)$\\
G109.870$+2.114$ (-10.5)  &   $160(110-310)$     & $9.0(8.8-9.8)$     &$6.50(5.75-7.80)$   &$-5.5(-9.0 - -5.5)$  &$90(10-100)$\\
G109.870$+2.114$ (-5.0)   &   $220(195-270)$     &$13.7(12.4-13.8)$   &$5.75(5.50-9.00)$   &$-7.0(-7.3 - -5.9)$  &$10(10-30)$ \\
G111.542$+0.776$          &   $190(150-290)$     &$12.1(12.0-12.7)$   &$5.75( 5.50-7.00)$  &$-7.5(-7.8 - -6.7)$  &$60(20-70)$\\
G111.542$+0.776$(-56.4)  &    $480(70-600)$      &$8.4(8.3-8.8)$      &$3.00(3.00-6.50)$   &$-5.5(-6.5 - -5.5)$  &$100(10-100)$\\
G111.542$+0.776$(-59.0)  &    $320(280-600)$     &$14.0(13.5-14.0)$   &$6.00(5.50-7.00)$   &$-5.5(-8.5 - -5.5)$  &$30(10-50)$\\
G133.947$+1.064$          &   $150(140-160)$     &$12.0(11.9-12.1)$   &$5.50(5.30-6.50)$   &$-7.5(-7.8 - -7.1)$  &$100(90-100)$\\
G188.946$+0.886$          &   $130(35-600)$      & $9.0(8.4-9.7)$     &$5.75(3.00-9.00)$   &$-9.0(-9.0 - -5.5)$  &$60(10-100)$\\
G192.600$-0.048$          &   $220(210-240)$     &$12.4(12.3-12.5)$   &$6.00(5.30-6.70)$    &$-7.0(-7.4 - -6.6)$  &$10(10-15)$\\
\hline
\end{tabular}
\end{table*}

\section{Discussion}

\subsection{Methanol emission in G192.600--0.048  by SMA observations}
 
G192.600$-0.048$ (S255IR) is the first and so far the only source where the maser emission in the  line belonging to the $J_1 - J_0$ A$^{-+}$ line series has been detected \citep{Zinchenko_et_al_2017_AA_606}. 
This maser line  was discovered in high angular resolution observations with ALMA in 2016  when it had the  peak flux density of about 25~Jy.
In the later ALMA observations in 2017 its peak flux density decayed by about 40\% \citep{Liu2018}. Substantial velocity gradient in the line emission was observed.
As a result, the line spectrum of the brightest position in high resolution observations peaks at a velocity of about 2.3~\kms\ while the spectrum of the line emission integrated over the whole emitting region with the size comparable to SMA beam peaks at a different velocity of $\sim$4~\kms. 

\begin{figure}
	\includegraphics[width=\columnwidth]{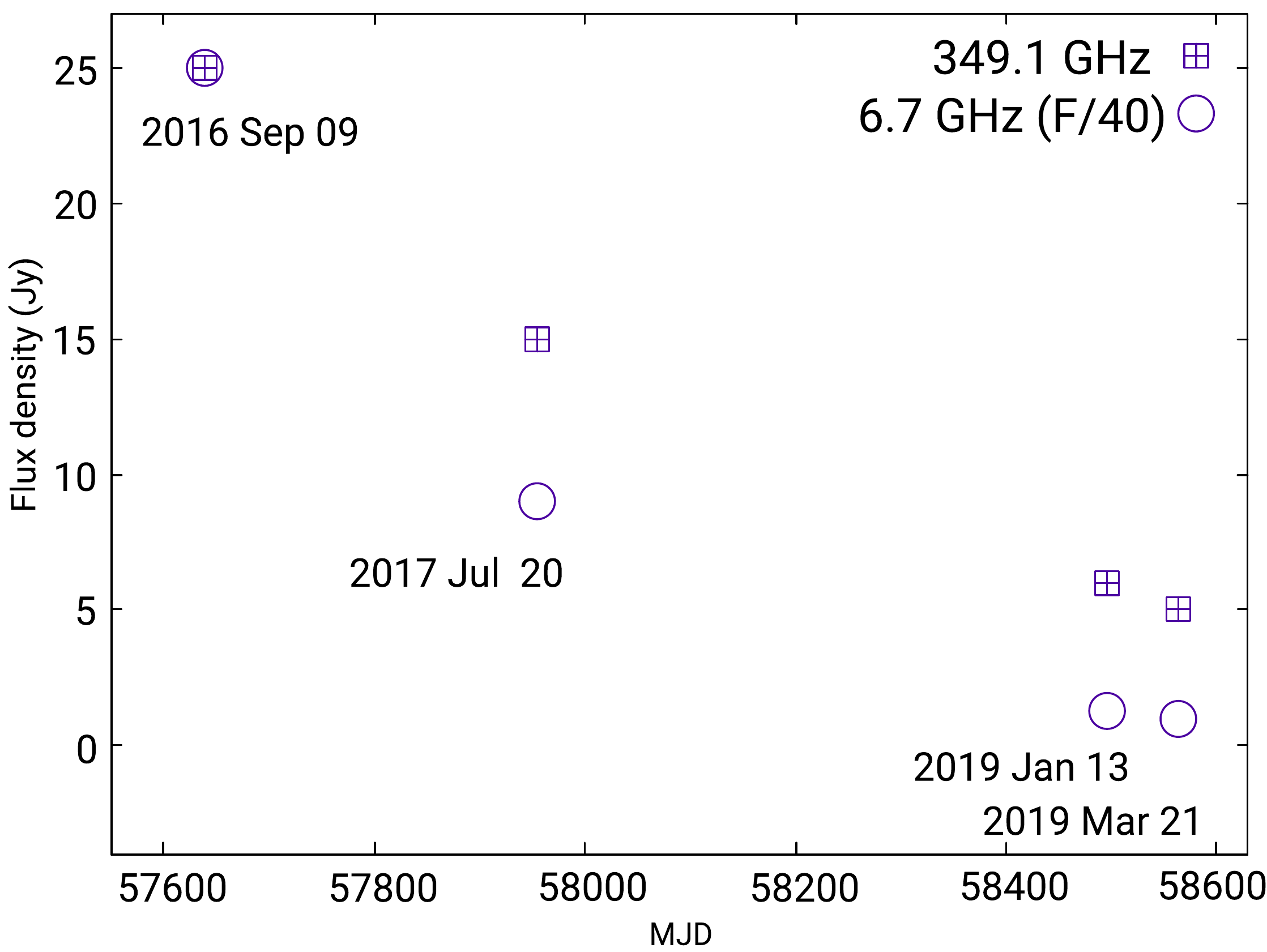}
    \caption{ Flux densities of the 349.107 GHz line in G192.60-0.05 (S255IR) which were observed in 2016 September \citep{Zinchenko_et_al_2017_AA_606} and 2017 July \citep{Liu2018} with ALMA, in 2019 January with IRAM 30m, in 2019 March with SMA (squares) and the 6.7 GHz maser flux densities   of the brightest in 2016 component at 6.3~\kms \citep{2017_M} scaled by a factor of 1/40 on the same dates (circles). The flux densities on the other dates are based on the RT-32 monitoring results.  Calibration uncertainties do not exceed the symbol size.}
    \label{fig:14_1_14_0_by_MJD}
\end{figure}

The data presented here show that the flux density in this line had the value  about 5~Jy (SMA) and 6~Jy (IRAM-30m)    (Fig. \ref{fig:14_1_14_0_by_MJD}). These values are consistent with each other within the calibration uncertainties and indicate a further decay of the maser emission to practically zero level since the line profile does not differ from the profiles of the other presumably thermal lines.

 One can see (Fig. \ref{fig:14_1_14_0_by_MJD}) that such a decay happens   contemporaneously  with the decay of the main component of the 6.7 GHz line. 
According to models presented by \citet{Cragg_et_al_2005_MNRAS_360}, the drop of the external dust temperature below about 150~K  greatly  
  reduces 
the 6.7~GHz maser line intensities. It could be the same reason for the decay of the emission in the $14_1 - 14_0$ A$^{-+}$ transition at 349.1~GHz.

\begin{figure}
   \includegraphics[width=\columnwidth]{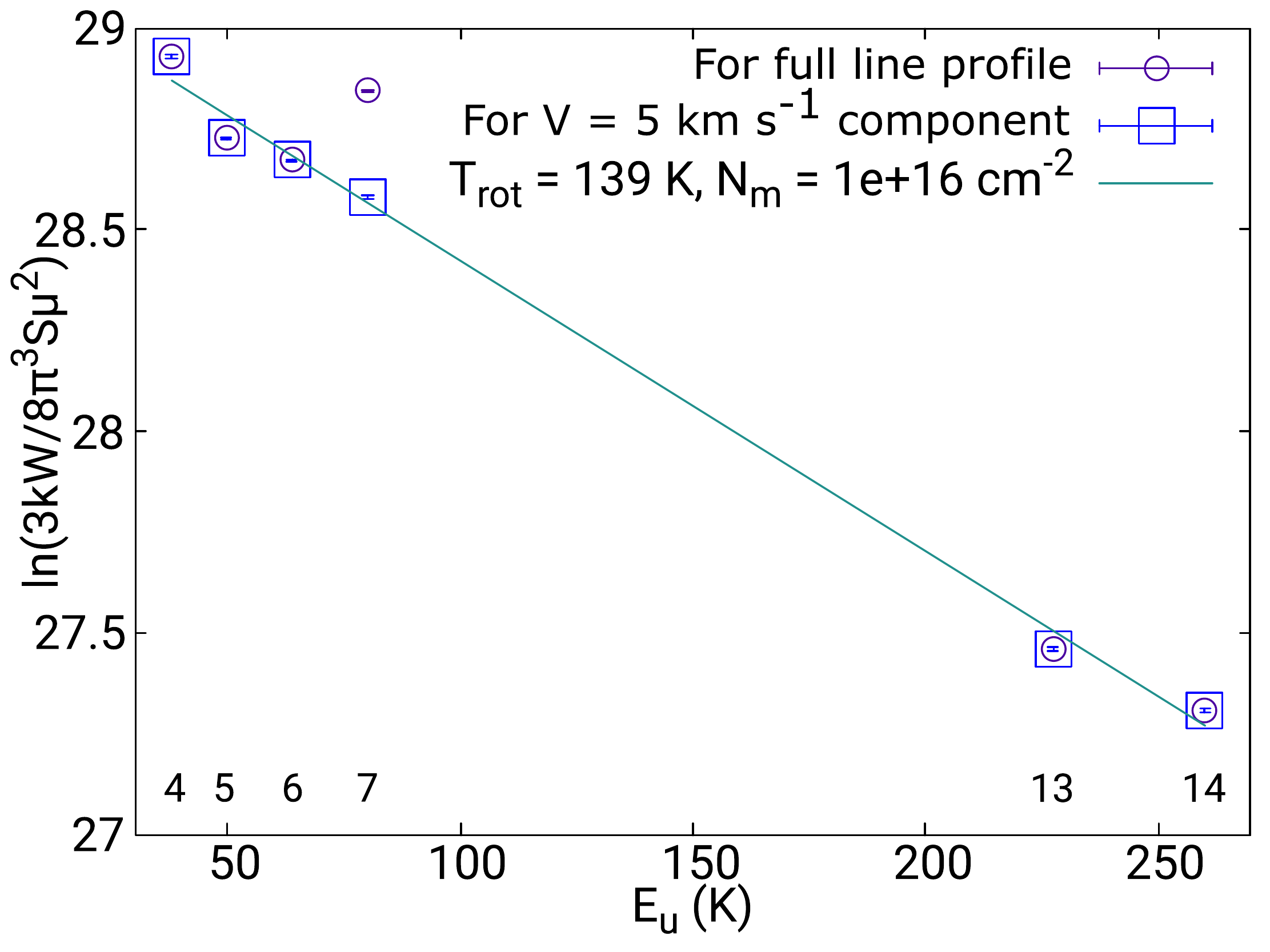}
    \caption{ Rotational diagram for the transitions in the $J_1 - J_0$ A$^{-+}$ series with $J=4-7, 13,14$ that were observed with SMA toward G192.60-0.05 (S255IR). For all 6 transitions, data points calculated with the full line profile are marked by circles and those calculated only for the velocity component at 5~\kms\ are marked by squares. The quantum numbers $J$ are labeled at the bottom. 
    }     
    \label{fig:J1_J0Amp_rot_dia}
\end{figure}

 At the same time, our SMA data show the brightest emission in the $7_1-7_0$ A$^{-+}$ line at 314.9~GHz (Fig.~\ref{fig:J1_J0_Amp_sp_SMA}, upper panel). In the rotational diagram  (Fig.~\ref{fig:J1_J0Amp_rot_dia})  one can see that the total flux density (circle) for
the $J=7$ line has an excess in emission.
When fitting the spectral profile of the $7_1-7_0$ A$^{-+}$ line with two Gaussian, we see an additional bright component at an velocity of about 2.8~\kms\ (Fig.~\ref{fig:J1_J0_Amp_sp_SMA}). This additional component is close   in  velocity to the maser component reported earlier by \citet{Zinchenko_et_al_2017_AA_606}. This component is clearly seen in the $7_1-7_0$ A$^{-+}$ line and   can  be  tentatively   recognized  in the   profile of the  $6_1-6_0$ A$^{-+}$ line at 311.9~GHz.   In the other lines of the series this component is not   seen   (Fig.~\ref{fig:J1_J0_Amp_sp_SMA}).  

As mentioned in Section~\ref{sec:sma_res} the 2.8~\kms\ component of the $7_1-7_0$ A$^{-+}$ line is observed in an extended region and most probably arises in the rotating disk-like structure around the massive protostar.

Since the emission in the 2.8~\kms component of the $7_1-7_0$ A$^{-+}$ spectral profile  t coincides in velocity with the maser component of   
the $14_1 - 14_0$ A$^{-+}$ line  observed   in 2016 \citep{Zinchenko_et_al_2017_AA_606},
 we propose that the spectral component at 2.8~\kms of the $7_1 - 7_0$ A$^{-+}$ transition   also has the maser nature.
In Section~\ref{sec:maser_conditions_J1_J0Amp} we   have shown   that under certain conditions the maser occurrence for the transitions of the $J_1 - J_0$ A$^{-+}$ series is possible. 

We see that in 2019 the maser   component   detected in 2016 in the $14_1 - 14_0$ A$^{-+}$ transition has practically disappeared, while the $7_1 - 7_0$ A$^{-+}$ transition probably shows the maser amplification. We tried to search for a set of physical parameters, which can explain such a situation.
We found out that  
an inverted population in the transition $14_1 - 14_0$ A$^{-+}$ at 349 GHz requires temperatures above 170~K, specific column densities from $10^{9}$~cm$^{-3}$s to~$2.5 \cdot 10^{11}$~cm$^{-3}$s and a methanol fractional abundance above $3.16 \cdot 10^{-7}$ (see Tab.~\ref{tab:J1_J0Amp_mod_res}). On the other hand, the transition $7_1-7_0$ A$^{-+}$ at 314 GHz can be inverted at temperatures from 140~K and with specific column densities from $3.2 \cdot 10^{8}$~cm$^{-3}$s to $2.5 \cdot 10^{11}$~cm$^{-3}$s.  
Thus, the situation when the transition $7_1-7_0$ A$^{-+}$ is inverted while the $14_1 - 14_0$ A$^{-+}$ transitions  is not masering can be realized at temperatures from 140 to 170 K.
 It is  noteworthy  that the values of the   temperature   and the methanol column density   estimated from  the rotational diagram (see Fig.~\ref{fig:J1_J0Amp_rot_dia})  are consistent with this assessment.

Thus we can conclude that it is very likely to be the case after accretion burst in S255IR that the dust cools down much more quickly than the gas. 
It is also necessary to mention that the accretion burst has a transient nature and non-stationarity effects can play important role in this case for the masers with the radiative source of pumping \citep{2018IAUS..336...59M, 2012IAUS..287...13S}. Both rise and drop of the maser emission in S255IR can be considerable on the time scales of days \citep[figures 1 and 2 in][]{2018A&A...617A..80S} which is comparable to the timescale of relaxation for Class~II methanol masers with their lengthy and staggery pump cycles \citep{1994ApJ...433..719S}.
 
\begin{figure}
	\includegraphics[width=\columnwidth]{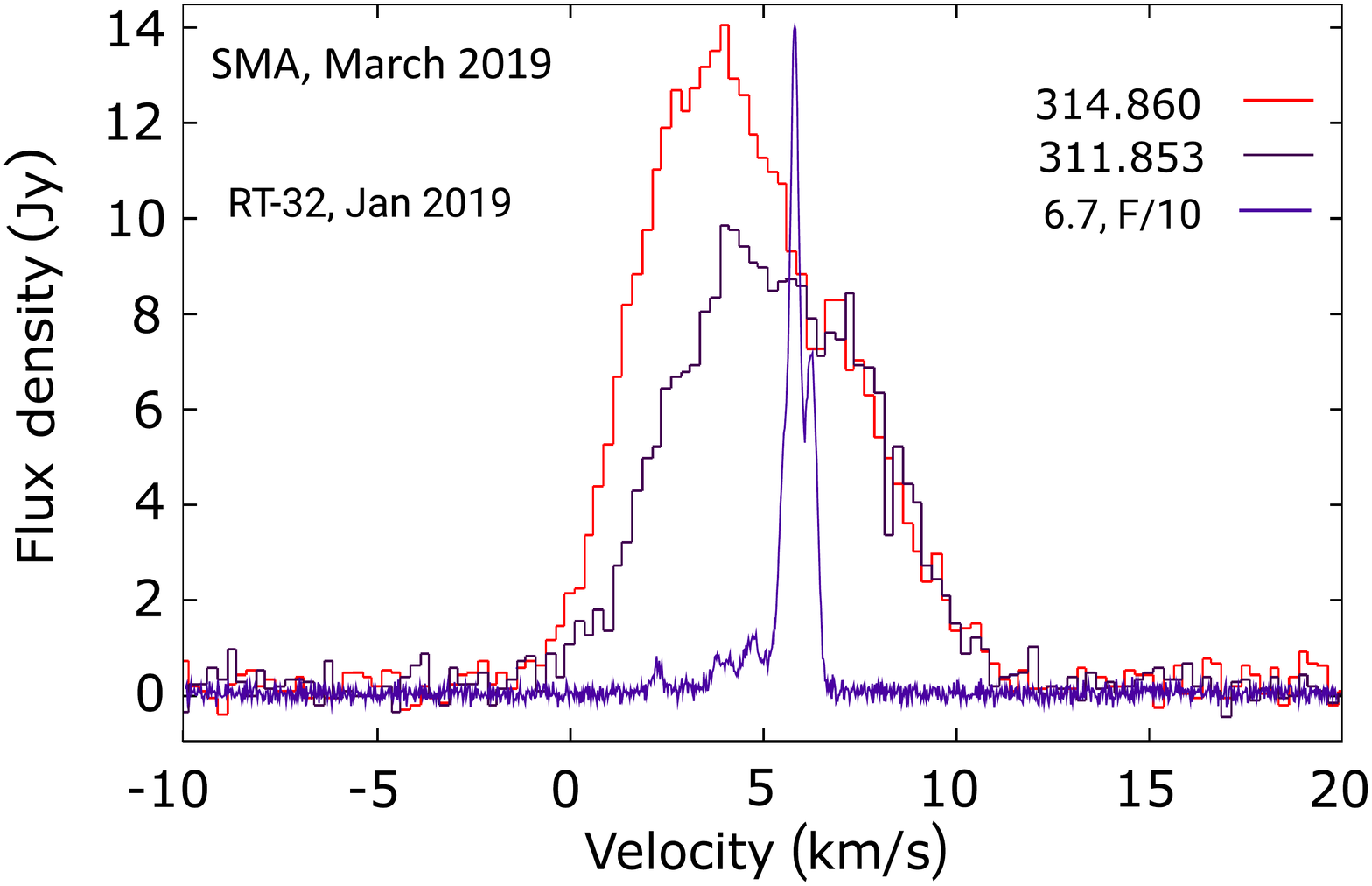}
    \caption{  An overlay of the spectra of the 314.860~GHz and 311.853~GHz lines observed with the SMA toward 
    G192.60-0.05 (S255IR) and the 6.7 GHz Class II methanol maser spectrum observed with RT-32 (VIRAC) 
    toward this source.}
    \label{fig:G192_6p7_314_311_GHz}
\end{figure}

Finally, it is worth mentioning that the maser emission at 6.7~GHz spans  from  about 1 to 7~\kms. 
There are 4 weak peaks at the velocities from $\sim 2.3$ to 4.8~\kms at 6.7~GHz  
which  corresponds to the suspected maser emission in the lines of the $J_1 - J_0$ A$^{-+}$ series (see Fig.~\ref{fig:G192_6p7_314_311_GHz}). At the same time, the main peak at 6.7~GHz is observed at a very different velocity (5.9~\kms).   It  shows that the conditions for the formation of the maser emission at 6.7~GHz and in the lines of the $J_1 - J_0$ A$^{-+}$ series are quite different. This   is reflected  by our modelling estimations of parameters for inversion of populations of the $J_1 - J_0$ A$^{-+}$ series levels and 6.7~GHz transition levels.

\subsection{Survey with IRAM-30m}

Rotational diagrams display rather large scatter of points exceeding observational uncertainties for most of the sources (Fig.~\ref{fig:rot_dia}). 
It suggests that the excitation of the methanol transitions shows considerable deviations from LTE. So, for determining the physical parameters we used non-LTE LVG approach and found the models which realize minimum $\chi^2$ deviation from the observed values (Table~\ref{tab:phis_params_res}). 
Unfortunately, the uncertainties of some estimates are very large. 
This may be caused by the small number of methanol lines that were confidently registered, for example, toward the sources G008.831$-0.028$, G035.200$-1.736$ and G188.946$+0.886$. On the other hand, large uncertainties can be caused by inhomogeneity of the source within the 7{\farcs}5 beam, which is likely true for the most of the sample sources. For a more accurate estimation of the physical parameters, observations with a higher angular resolution are required.

\begin{figure*}
    \includegraphics[width=0.95\linewidth]{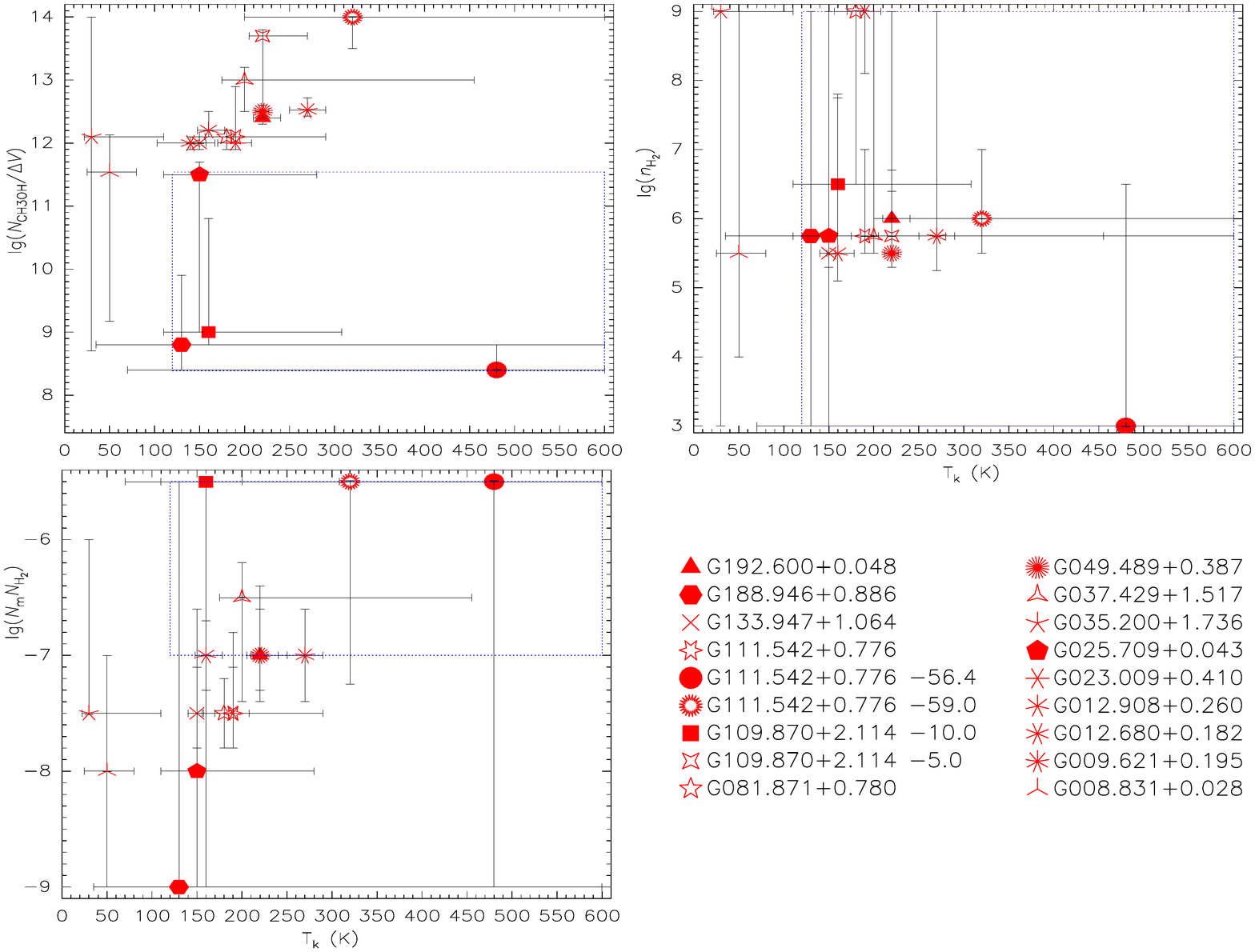}
    \caption{  The best-fit parameters inferred from the LVG analysis, including specific methanol column density (lg($N_{M}/\Delta V$)), number density (lg($n_{H_2}$)), methanol fractional abundance (lg($N_{M}/N_{H_2}$)) and kinetic temperature $T_{k}$, are shown in different combinations for the 15 sources observed with IRAM 30m. Ranges of parameters where inversion of populations can occur are delineated by the dashed lines. 
}  \label{fig:param_Tk}
\end{figure*}

Nevertheless, we may conclude that  all of the sources, except G008.831$-0.028$ and G035.200$-1.736$, have $T_{k}$ exceeding 100~K.  
These temperatures are within the range where inversion of population in the $J_1 - J_0$ A$^{-+}$ series levels can occur (Fig.~\ref{fig:param_Tk}). 
 
All of the sources in one Gaussian fit have number densities exceeding $3\cdot10^5$~cm$^{-3}$. For the sources G023.009$-0.410$ and G081.871+0.780 we find their number densities exceeding $3\cdot10^6$ and $10^8$~cm$^{-3}$, respectively, which in combination with the temperatures of 190 and 180 K correspond to hot cores.

The fractional abundances for most of the sources vary from $10^{-8}$ to  $3\cdot10^{-7}$.   
These methanol fractional abundances are higher than the values typically seen in the dark molecular clouds ($\sim 10^{-9}$), and  can be  attributed to thermal evaporation of dust grain icy mantles. Temperatures above $\sim 80$~K lead to evaporation of methanol from the dust grain mantles. Therefore, it is possible that most of the observed sources contain internal heating sources.

For the sources G192.600$-0.048$, G037.429$+1.517$ and G012.680$-0.182$ the beam filling factor is about 10–-20\%  with confidence 68\%.  This means that these sources are very inhomogeneous within the 7{\farcs}5 beam and probably consist of clumps with smaller sizes. Thus, the densities of the clumps could be higher than the densities estimated above. On the other hand, for the sources G049.489-0.387, G081.871+0.780 and G133.947+1.064 a beam filling factor of 80--100\% was estimated. 

As can be seen in Fig.~\ref{fig:param_Tk}, only the velocity component at $-56.4$~\kms in G111.542+0.776 has all its physical parameters falling within the ranges within which the population inversion of the $J_1-J_0 A^{-+}$ series is allowed.  For this component, we see the inversion of the level populations of the $J_1 - J_0$ A$^{-+}$ series of transitions in the modelled brightness temperatures. The physical parameters for this component differ from the others in its highest fractional abundance  $3\cdot10^{-6}$, its lowest methanol specific column density $3\cdot10^{8}$~cm$^{-3}$s,  and a high temperature of 480 K.  The hydrogen number density in this model was estimated with a large uncertainty as $< 3\cdot10^{6}$~cm$^{-3}$. 
This source G111.542+0.776 is discussed in more detail in  Section~\ref{sec:ouG111p54}

We therefore conclude that according to our model the population inversion in the $J_1-J_0 A^{-+}$ series transitions is expected to be rare in the type of objects from our sample. This is consistent with our observations, in which the sources do not show obvious maser features.
 
\subsubsection{Emission in the source G109.870+2.114}
There are spectral components at two distinct velocities: about $-5$ and $-10$~\kms \,  in the methanol line profiles toward Cep~A  (Fig.~\ref{fig:all_sou_sp}). One can see that the spectra of the lines from the $J_1-J_0 A^{-+}$ series have almost equal peaks at both velocities, though in the $J=11$ line  at 331.502~GHz the component at $-10$~\kms\ is brighter. At the same time   the  emission at $-10$~\kms\ in the other considered methanol lines is obviously weaker compared to the emission at $-5$~\kms.

The  emission at $\sim -5$~\kms\ is associated with the brightest radio continuum source in this region, HW2, where \citet{2005Natur.437..109P} detected the presence of a flattened disk-like structure. The emission from the Class II methanol maser at 6.7~GHz is associated with the HW2 disk and there is no emission at the other velocity for this transition. 

The  spectra with profiles similar to those of the lines from the $J_1 - J_0$ A$^{-+}$ series, with two components at about $-5$ and $-10$~\kms,  were observed by \citet{2007ApJ...660L.133B} with   a  resolution  of  $1-2''$ in a number of molecular lines.   The  authors concluded that some species exhibit a main peak of emission   at only  one of the two distinct velocities. For example CH$_3$OH, NH$_2$CHO, and H$_2$CS emit predominantly at $-5$~\kms\ but HC$_3$N, SO$_2$ emit at  $-10$~\kms. Meanwhile, they note that "a few abundant high-density tracers like CH$_3$CN and C$^{34}$S show emission of nearly equal strength toward both positions" \citep{2007ApJ...660L.133B}. Apart from   the  kinematic dichotomy, the authors reported   a  thermal differentiation with temperature about 120 K for the object emitting at velocity $-5$~\kms and  230--310~K for the   other  object emitting at  $-10$~\kms. 

According to \citet{2009ApJ...703L.157J} and the references therein, the object   associated with  the emission at $-10$~\kms is likely the powering source of the small-scale SiO outflow and hosts a massive protostar.   These  authors proposed that since this object has not yet ionized its surroundings, it is at an earlier stage of evolution than the HW2 source.  Consequently, we could conclude that  the  lines of the series are tracing a hot region around the protostar. Another possibility is that the emission at $-10$~\kms\ is formed in the HW3d source, which is also associated with  a high-mass young stellar object \citep{2012ApJ...748..146C}. This has to be  examined with high angular resolution observations.

\subsubsection{Spectral peculiarity in the source G111.542+0.776}\label{sec:ouG111p54}

In the spectra of the lines of the $J_1 - J_0$ A$^{-+}$ series from G111.542+0.776 (NGC 7538C), one can see clearly non-Gaussian profiles while for other bright lines the spectral profiles are Gaussian (see Fig.\ref{fig:all_sou_sp}). Apart from the main spectral component at $V_{lsr}$ $-59$~\kms, a relatively weak spectral component at the $V_{lsr}$ about $-56$~\kms can be distinguished. It is unlikely an interloper from some other molecule since the same velocity component are present in three lines with different frequencies and there are no corresponding components in the other sources. 

\begin{figure}
	\includegraphics[width=\columnwidth]{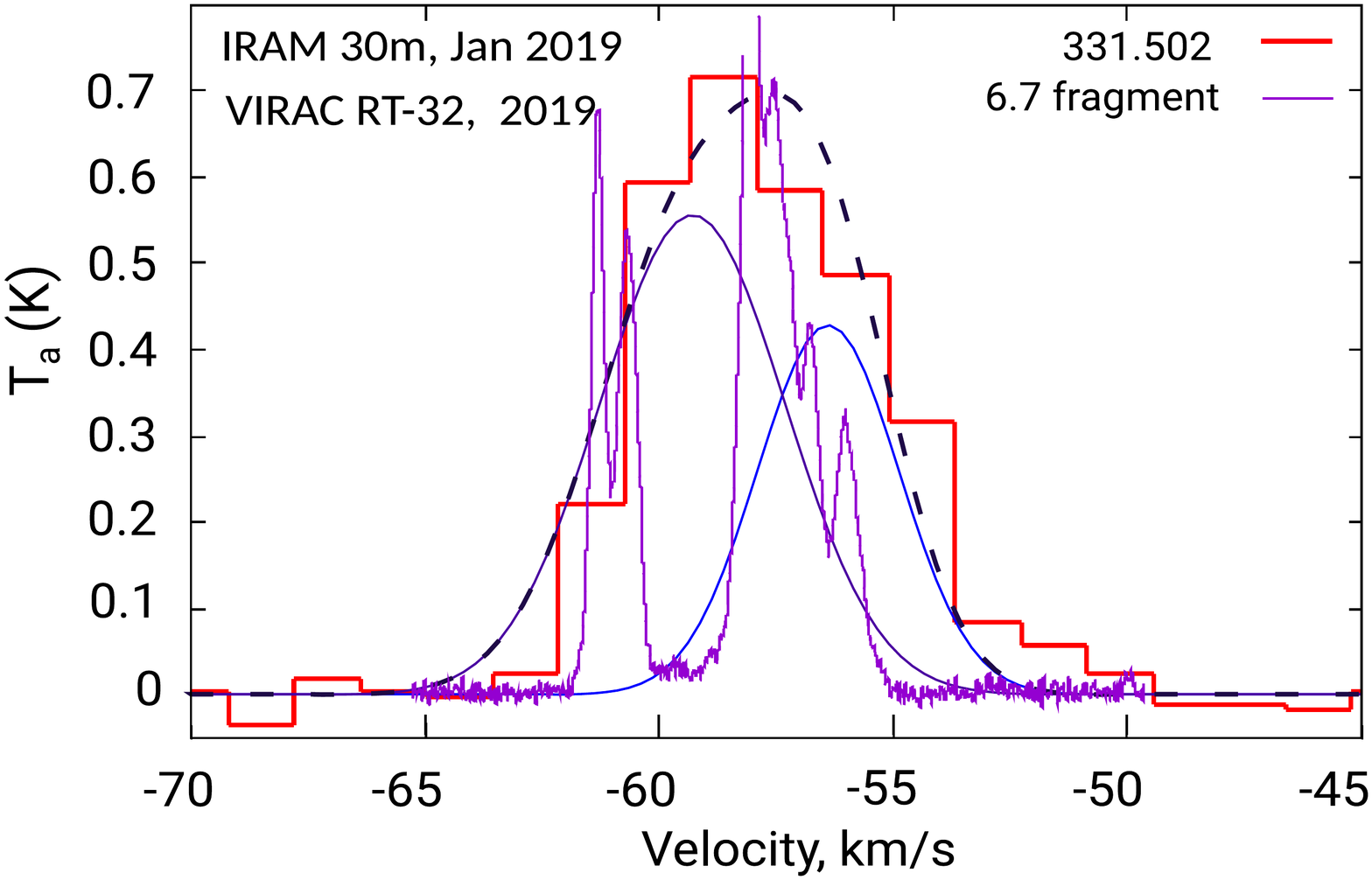}
    \caption{An overlay of the 6.7 GHz spectra (plotted in blue) and the 331.505 GHz (the brightest in the $J_1 - J_0$ A$^{-+}$ series) spectra (plotted in red) in G111.54+0.77, which were observed roughly simultaneously in Jan. 2019.}
    \label{fig:G111p54_6p7_331p505GHz}
\end{figure}

The component at $V_{lsr}$ about $-56$~\kms\ was the brightest component in the 6.7~GHz maser spectra observed in 1991 and 1999 \citep{1991ApJ...380L..75M,2000A&AS..143..269S}.  According  to the observations in 2012 by \cite{Hu_et_al_2016}, this spectral component has an intermediate brightness among the other components at 6.7 GHz. If we overlay the spectra of the brightest line from the $J_1 - J_0$ A$^{-+}$ series at 331.505~GHz with the maser emission at 6.7~GHz observed in January 2019 (see Fig.~\ref{fig:G111p54_6p7_331p505GHz}), we can see that   the additional  spectral component at $-56.4$~\kms\ at 331.5~GHz coincides with   the  two relatively weak maser components at $-56.8$ and $-56.0$~\kms\ at 6.7 GHz. \citet{2009A&A...501..999P} detected  maser emission  in the 6.7 GHz and 12 GHz methanol lines at these velocities and interpreted it as the emission from an edge-on disk. \citet{2013A&A...558A..81B} indicated the presence of active star-formation processes in the form of fragmentation, infall, and outflows in the object. Moreover they report two distinct bright components in the spectrum of the $15_1 - 15_0$ A$^{-+}$ methanol line at 356 GHz (see Fig. 3 from \citet{2013A&A...558A..81B}).  This line belongs to the series under consideration.  The authors noted  that the $15_1 - 15_0$ A$^{-+}$ line ``is among the few lines that do not show any absorption at high spatial resolution''  and said that it looks like maser emission or could be emission tracing a circumstellar disk  with a high excitation temperature.  It is noteworthy that the velocities of the components in the 356~GHz line spectrum from \citet{2013A&A...558A..81B} are in general agreement with the line velocities reported here.  In addition,  \citet{2015A&A...573A.108G} noted that there are  some indications of a rotating core with two small disks in the source at the velocities under consideration.  

In order to distinguish the emission of different components, we fitted all observed methanol lines with two Gaussian having fixed  velocity values of $-59$~\kms\ and $-56.4$~\kms, and a fixed line width of 3.5~\kms\ for the latter component. We do not consider the emission from blends at 346.20272~GHz and 346.20427~GHz since the velocity difference between them is comparable to that between components, so distinguishing components in these blended lines is impossible.

Using  the  methanol line intensities that were determined   from the two-Gaussian   fit, 
we estimated the physical parameters of these components (corresponding to the $\chi^2$ minimum) and their confidence intervals (Table~\ref{tab:phis_params_res}). 

By the LTE analysis of the NH$_3$ emission, \citet{2015A&A...573A.108G} suggested that the molecular gas in the core has a temperature of 280 K,  with a potential hotter component up to 500~K that coincides with our estimates in Table~\ref{tab:phis_params_res}. 

Remarkably, that populations of the $J_1 - J_0$ A$^{-+}$ series levels are inverted in the model for the component at $V_{lsr}=-56.4$~\kms. In contrast, in the model for the component at $V_{lsr}=-59$~\kms\ all transitions have quasi-thermal excitation. Despite the similar temperatures (about 500 K), the  best fit models have a large difference in the specific column density and number density ($2.5\cdot10^9$ cm$^{-3}$s and $10^3$ cm$^{-3}$ for the component at $V_{lsr}=-56.4$~\kms, while $10^{14}$ cm$^{-3}$s and $10^6$ cm$^{-3}$ for the component at $V_{lsr}=-56.4$~\kms).
At low densities the populations of the energy levels  usually do not follow the Boltzmann distribution, so it is easier to achieve   great deviation from LTE in the form of population inversion. Furthermore, to produce the inversion of level populations we need sufficiently high column density, however, according to our model the specific column density should not exceed $\sim4\cdot10^9$ cm$^{-3}$s (Table~\ref{tab:J1_J0Amp_mod_res}). 

The modeling results and observational data presented above   indicate that the component  observed in the $J_1-J_0$ A$^{-+}$ series at $V_{lsr}=-56.4$ ~\kms\ may well be a weak maser.

\subsection{The $J_1-J_0$ A$^{-+}$ methanol line series as a tracer of the physical conditions in star-forming regions}

Our observational data and theoretical modeling show that masering is possible in the $J_1-J_0$ A$^{-+}$ methanol line series under certain conditions. These conditions imply rather high gas temperature  ($> 110$~K). The required temperatures   are  different for different transitions within the series (from $>110$~K for transition with $J = 1$ to $>210$~K for transition with $J = 19$),    which makes it possible to constrain the temperature from observations of a set of lines.   
 
The case of S255IR shows that the maser effect in these lines can be an  important  tracer of luminosity bursts in high mass star-forming regions. In such cases the temporal variations of the physical parameters can be investigated. The model used here is rather simple. A more advanced approach  has to be used to achieve better estimates of the physical parameters. 

The maser effect in the lines of the $J_1-J_0$ A$^{-+}$ methanol line series is rather rare in our sample, which   is composed  of a set of the brightest 6.7~GHz methanol maser sources. However, it is worth noting that the physical conditions required for the maser effect in these lines are apparently significantly different from those for the 6.7~GHz methanol masers. It is   possible that  the $J_1-J_0$ A$^{-+}$ methanol masers are more frequent in other types of astronomical objects   such as  hot cores.

In general, the maser effect in these lines is rather weak in this sample and hardly can be discovered in single-dish observations. However the existing submillimeter interferometers are well suited for this purpose. 

\section{Conclusions}

Motivated by the recent detection of the unpredicted maser emission in the $14_1 - 14_0$ A$^{-+}$ line toward the high-mass star-forming core S255IR-SMA1 \citep{Zinchenko_et_al_2017_AA_606}, where the disk-mediated accretion burst happened in 2015, we performed an observational and theoretical study of the maser effect in the $J_1-J_0$ A$^{-+}$ methanol lines.

1. The maser emission in the $14_1 - 14_0$ A$^{-+}$ line at 349.1~GHz toward S255IR-SMA1 detected in 2016, decayed to a zero level in early 2019. At the same time we almost certainly detected a maser component in the $7_1-7_0$ A$^{-+}$ line at 314.9~GHz and probably in some other lines of this series toward this object. A simple theoretical model shows that the corresponding change of the brightest maser line of the series
can be explained by the temperature decrease in S255IR-SMA1 since 2016. 
At the same time our current modelling does not explain the observed line ratios and maser line intensities. This is possibly a result of the model simplicity and non-stationarity effects in the maser pumping.

2. Up to 20 methanol lines were detected in a sample of the brightest 6.7~GHz methanol maser sources at the frequencies from 326.7 to 350.2 GHz in the observations with the IRAM 30m telescope. There is no obvious (bright) maser emission in the lines of the $J_1-J_0$ A$^{-+}$ series. However the emission in the lines of this series is relatively bright in all sources. In some of these sources there are spectral components which probably represent weak masers.
In particular, an additional spectral component at $V_{lsr} \sim -56$~km$^{-1}$ in the $J_1-J_0$ A$^{-+}$ line series   in the source G111.54+0.77 (NGC 7538C) was detected. The best fit model has inverted level populations for transitions of the $J_1-J_0$ A$^{-+}$ series.  

3. The maser effect in the lines of the $J_1-J_0$ A$^{-+}$ series is rare and is probably observed in quite hot and probably dense environments. Theoretical modeling supports this view. It does show inverted populations for these lines under such conditions. Therefore, maser emission in these lines may accompany luminosity flare events, like that in S255IR in 2015, and can probably serve as an indicator of the flare events.

\section*{Data availability}
The data used in this paper are available by request from the corresponding author.

\section*{Acknowledgements}

We are thankful to Sergey Parfenov for discussions of the methanol maser activity   and the  referee for comments and suggestions which helped us to clarify our statements and improve the presentation.
This work is based on observations carried out under project number 155-18  with the IRAM 30m telescope. IRAM is supported by INSU/CNRS (France), MPG (Germany) and IGN (Spain).
SVS and AMS were supported by the Russian Ministry of Science and Higher Education, No. FEUZ-2020-0030 for the work on the modeling of the CH$_3$OH emission using rotational diagram method. 
This work was supported by the Russian Science Foundation grants No. 17-12-01256 (the preparation of the observations, the SMA data analysis and the general discussion) 
and 
No. 18-12-00193-P (modeling of the CH$_3$OH excitation using LVG approach and the observations at the 30-m IRAM telescope). 
The initial IRAM data reduction and IRAM data analysis were supported by the RFBR grant 20-52-53054 and IRAM. 
The maser excitation analysis by AMS was supported by the BASIS Foundation.
SY Liu acknowledges the support from Ministry of Science and Technology through the grant MOST 109-2112-M001 -026-.
A. Aberfelds acknowledges the support from the Latvian Council of Science Project “Research of Galactic Masers” Nr.: lzp-2018/1-0291.




\bibliographystyle{mnras}
\bibliography{salii_et_al_2020} 








\bsp	
\label{lastpage}
\end{document}